\documentclass[usenatbib]{mnras}
\usepackage{amsmath, amssymb}
\usepackage{graphicx}
\usepackage{url}
\usepackage{hyperref}
\usepackage{multirow}
\usepackage{tabularx}
\usepackage{makecell}

\title{CSS1603+19: a low-mass polar near the cataclysmic variable period minimum}
\author[Y. Liu et al.]{Yiqi Liu,$^{1}$\thanks{Email: andrew.liu@jhu.edu}
Hsiang-Chih Hwang,$^{2}$
Nadia L. Zakamska$^{1,2}$
and John R. Thorstensen $^{3}$
\\
$^{1}$Department of Physics \& Astronomy, Johns Hopkins University, 3400 N Charles St, Baltimore, MD 21218, USA\\
$^{2}$School of Natural Sciences, Institute for Advanced Study, Princeton, 1 Einstein Drive, NJ 08540, USA\\
$^{3}$Department of Physics and Astronomy, Dartmouth College, Hanover, NH 03755, USA
}
\date{Accepted 2023 April 16. Received 2023 April 16; in original form 2022 November 28}

\graphicspath{{figures/}}

\begin{document}
\maketitle
\label{firstpage}
\pagerange{\pageref{firstpage}--\pageref{lastpage}}
\begin{abstract}
    CSS1603+19 is a cataclysmic variable (CV) with an orbital period of 81.96 min, near the minimal period of cataclysmic variables. It is unusual in having a strong mid-infrared excess inconsistent with thermal 
    emission from a brown dwarf companion. Here we present time-resolved multi-wavelength observations of this system. WISE photometry indicates that the mid-infrared excess displays a one-magnitude eclipsing-like variability 
    during the orbit. We obtained near-infrared and optical spectroscopy using Gemini, MDM and APO telescopes. Near-infrared spectra show possible cyclotron features indicating that the white dwarf has a magnetic field 
    of about 5MG. Optical and near-infrared spectra display double-peaked emission lines, with both components showing strong radial velocity variations during the orbital period and with the broad component 
    leading the narrow component stably by about 0.2 of the orbital phase. 
    We construct a physical model informed by existing observations of the system and determine that one component likely originates from the accretion column onto the magnetized white dwarf in synchronous 
    rotation with the orbital motion and the other from the Roche overflow point. This allows us to constrain the masses of the binary components to be $M_1>0.24 M_{\odot}$ for the white dwarf accretor and 
    $M_2=0.0644\pm0.0074 M_\odot$ for the donor. We classify the system as an AM Herculis star, or a polar. It has likely completed its stint on the period gap, but has not yet gone through the period bounce. 
\end{abstract}

\begin{keywords}
    binaries: close -- binaries: spectroscopic -- novae, cataclysmic variables -- stars: low mass -- white dwarfs
\end{keywords}

\section{Introduction}
\label{sec:intro}

Cataclysmic variables (CVs) are compact binaries consisting of an accreting white dwarf and a Roche-lobe filling companion star, or `donor' \citep{Warner1995}. 
The formation of CVs is driven by angular momentum dissipation of a post-common-envelop binary, where the orbital period of the system gradually decreases from an initial orbital period P$_{\rm orb} \gtrsim 10$ hours to an orbital period minimum of
P$_{\rm min}\simeq 80$ min \citep{Patterson1984, Ritter2003, 2011Knigge_CV_evo, 2016Kalomeni_CV_evo}. 

The complex distribution of observed orbital periods of CVs is accounted for by the disrupted magnetic braking scenario 
\citep{Rappaport1983}. At $P_{\rm orb}\gtrsim 3$ hours, the angular momentum loss for CVs is dominated by magnetic braking \citep{Patterson1984}. In this scenario, the low-mass donor's magnetic field drives ionized stellar wind, which co-rotates with the star's field out to the 
Alfv\'enic limit, exerting a resistant torque on the donor star and dissipating the angular momentum. In the meanwhile, in CVs the donor star is tidally locked and is co-rotating with the system's orbit. Thus, the angular momentum loss due to the stellar wind propagates to the angular momentum loss of the entire CV system
\citep{1981Verbunt_MB, 1986Paradijs_MB, 1988Hameury_magnetic_braking, 1987Mestel_magnetic_braking}.

When the orbital period drops to $\sim 3$ hours,  the donor star becomes convective and shrinks to a size below its Roche radius, which disrupts the magnetic dynamo for sustaining the magnetic braking \citep{1989Taam_magnetic_braking}. 
Because the mass loss rate of the donor directly translates to the angular momentum loss rate,  the cessation of magnetic braking leads to a sudden decline of both these 
physical quantities. The accretion ceases and the system evolves as a detached binary, so for a while the orbital period shortens only due to gravitational radiation, until 
the donor star re-fills its shrunken Roche lobe
\citep{1983Spruit_CV_period_gap}. The re-attachment of the donor to the Roche lobe at around $P_{\rm orb}=2$ hours re-initiates the mass transfer, and the system re-emerges as an active CV. 
As a result, there is a paucity of CVs with periods between 2 hours and 3 hours, a phenomenon known as the period gap 
\citep{1982Rappaport_compact_binary_evo, 1984Verbunt_mass_transfer_and_period_gap, Ritter2003, 2006Knigge_Period_gap}. 

The orbital period continues to shorten for these re-ignited CVs via gravitational radiation and mass loss. The mass transfer eventually causes the donor to go 
out of thermal equilibrium due to the Kelvin-Helmholtz time scale growing faster than the mass-loss time scale. In this situation, further mass loss results in an increase in orbital period \citep{Paczynski1981}, and the period distribution of CVs displays a minimum at  
P$_{\rm min}\simeq 80$ min \citep{1993Kolb_cv_P_min, 1999Kolb_cv_P_min, Ritter2003, Gaensicke2009}. 

In this paper, we present observations and modeling of an enigmatic CV CSS J160346.08+193540.0 (hereafter CSS1603+19). This object came to our attention as a periodic infrared variable with a period of P$=81.96$ min -- near the theoretical minimum for CVs -- identified during our exploration of the time-variability data from the Wide-field Infrared Survey Explorer \citep[WISE;][]{WISE_Spec_2010} which included the search for eclipsing binaries and other periodic variables \citep{Hwang_2020_EBs, Petrosky_2021}. CSS1603+19 exhibits a strong mid-infrared excess which cannot be accounted for by a brown dwarf companion, and a mid-infrared periodic variability of one magnitude. The source was known previously as a candidate ``polar" or AM Herculis star from its photometric characteristics \citep{Oliveira2020} and optical spectroscopy, with the radial velocity variability of emission lines indicating the same orbital period as the infrared light curve 
\citep{Thorstensen_2017cv_period_note}. Polars are CVs in which the magnetic field is high enough to prevent the formation of an accretion disk, to channel the accreting flow into streams along the field lines and to therefore synchronize the rotation of the white dwarf with the orbit.

In Section \ref{sec:data}, we describe the existing survey and follow-up data of CSS1603+19. In Section \ref{sec:measurements} we explain our fitting procedures 
for photometric, spectroscopic and light curve data. We present the analysis of these measurements and a geometric model in Section \ref{sec:analysis}. 
We conclude in Section \ref{sec:conclusion}. Emission-line wavelengths are given in air. To describe long-term variation over an extended period (on timescales of months) we use the terminology of `high' vs `low' states, and we use `bright' vs `faint' for the variability over the 82-minute orbital period. 

\section{Data}
\label{sec:data}

In this section, we describe the data we obtained for CSS1603+19. Our data include both targeted follow-up observations, as well as archival and survey data for the source.

\begin{figure*}
    \centering
    \includegraphics[width=\linewidth]{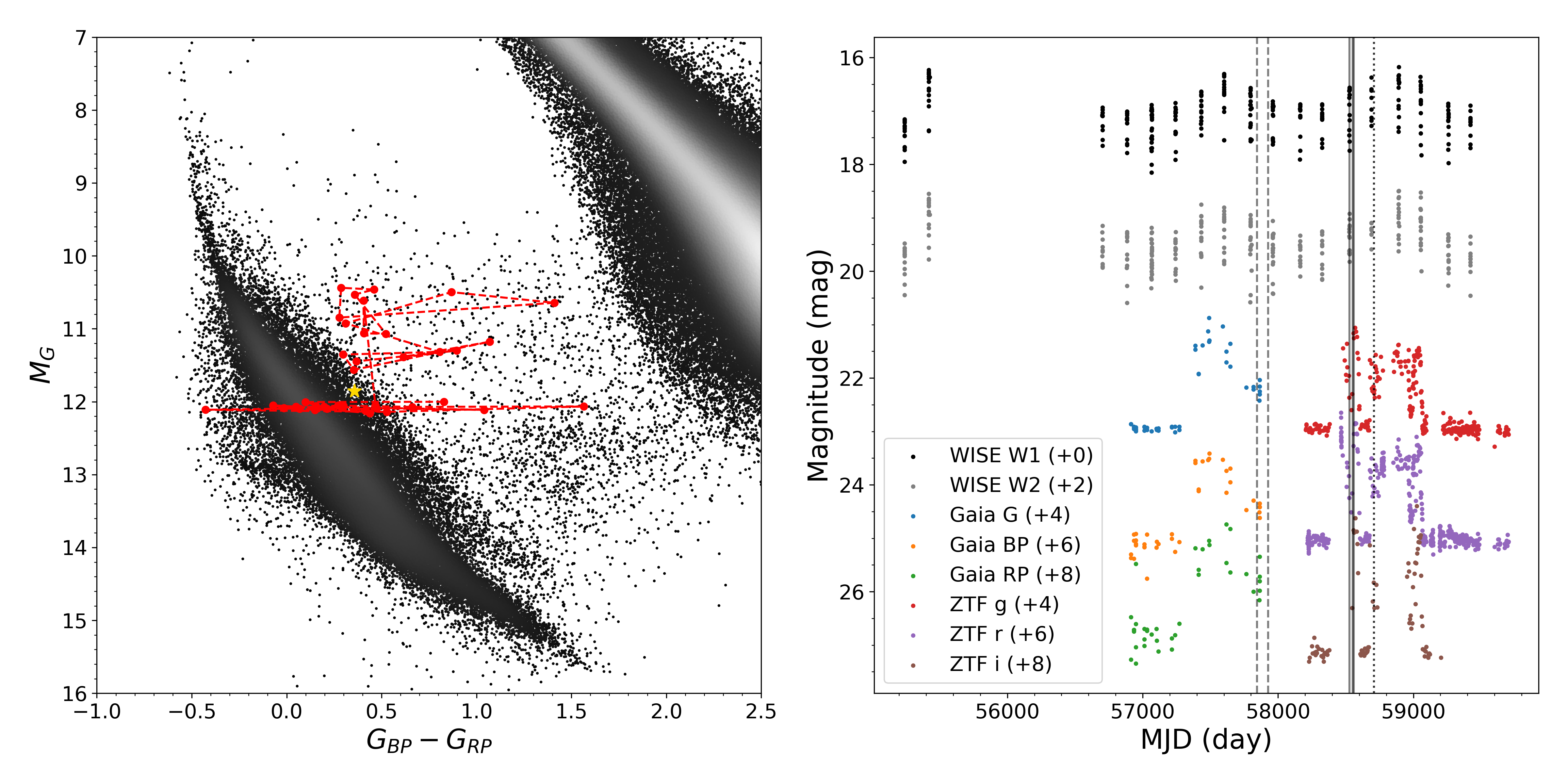}
    \caption{Left panel: {\it Gaia} color-magnitude diagram constructed using sources
    within 250 pc from the Sun. The gold-star marker shows the DR3 {\it Gaia}
    average measurement for CS1603+19. The red dots show the {\it Gaia}
    epoch photometry of CSS1603+19. Right panel: Raw photometry data, offset vertically for clarity as listed on the label. Dashed, dotted and solid lines delineate the epochs of optical MDM, optical APO and NIR GNIRS spectroscopy, correspondingly. }
    \label{fig:optical_raw}
\end{figure*}

\subsection{WISE Photometry}

CSS1603+19 is well detected by WISE in W1 (3.6 \micron) and W2 (4.5 \micron) with signal-to-nose ratios $>28$ in the AllWISE catalog, but is not detected in W3 (12 \micron) and W4 (22 \micron). CSS1603+19 was covered by both AllWISE and NEOWISE \citep{Mainzer_2011} eras of the mission. Thus, we were able to obtain reliable  mid-infrared light curves for $\sim 11$ years from 2010 to 2021, totaling
520 data points for single-epoch photometry.

We used W1 (3.6 \micron) and W2 (4.5 \micron) band data to construct the photometric light curve. The data were downloaded from NASA/IPAC Infrared Science Archive. To ensure the 
quality of single-epoch exposure, we followed the guideline provided by \citet{WISE_DATA_GUILD_2012} by adopting the criteria of 
\texttt{saa\_sep\,>\,0}, \texttt{moon\_masked\,=\,0}, \texttt{qi\_factor\,>\,0.9} for 
AllWISE\footnote{\url{https://wise2.ipac.caltech.edu/docs/release/allwise/expsup/sec3_1.html}}, 
and \texttt{saa\_sep\,>\,0}, \texttt{moon\_masked\,=\,0}, \texttt{qi\_factor\,>\,0.9}, \texttt{qual\_frame\,>\,0} for 
NEOWISE\footnote{\url{https://wise2.ipac.caltech.edu/docs/release/neowise/expsup/sec3_1a.html}}. Redundant exposures contained in 
AllWISE Multiepoch Photometry Database were excluded by matching source ID (\texttt{source\_id\_mf}) in the AllWISE and the NEOWISE catalogs\footnote{\url{https://wise2.ipac.caltech.edu/docs/release/allwise/expsup/sec3_2.html}}.

WISE photometry is measured in the Vega magnitude system. For consistency in this paper, we converted the W1 and W2 photometry to the AB-magnitude system
using the following equations\footnote{\url{https://wise2.ipac.caltech.edu/docs/release/allsky/expsup/sec4_4h.html}}:
\begin{align*}
    W1_{AB} = W1_{Vega} + 2.699, \\ 
    W2_{AB} = W2_{Vega} + 3.339.
\end{align*}

\subsection{ZTF Photometry}

We constructed the optical light curve of CSS1603+19 using data from Zwicky Transient Facility \citep[ZTF;][]{2019Masci_ZTF_Description, 2020ZTF_facility} at the 
Palomar Mountain in California. We use DR12 from ZTF, which provides observations of CSS1603+19 between March 2018 and May 2022 for over 4 years totaling 804 exposures. 
We used $g$, $r$, and $i$ bands for the light curves. The data are queried from NASA/IPAC Infrared Science Archive.
Quality filters are already applied to the light curve data produced by the archive.

\subsection{{\it Gaia} Photometry}

We supplemented our optical data using {\it Gaia} photometry from {\it Gaia} Data Release 3 \citep[DR3;][]{2016Gaia_mission, GaiaDR3Vallenari2022}. {\it Gaia} performed optical epoch photometry for 
CSS1603+19 from 2014 to 2017 totaling 128 data points from the G, BP, and RP bands. {\it Gaia} data are quality-filtered by the \texttt{rejected\_by\_photometry} and \texttt{rejected\_by\_variability}
flags.
The {\it Gaia} photometry supplements the optical photometry of ZTF observations.
The {\it Gaia} parallax of CSS1603+19 is $4.25\pm 0.12$ mas, giving the distance of the source from the Solar System to be $d = 235.3_{-6.4}^{+6.9}$ parsec. The coordinate of CSS1603+19 from {\it Gaia} DR3 is (ra, dec)=(240.94205667792, 	19.59439521850) in degrees at the reference epoch of J2016.0. 

In addition, we obtained $G$-band magnitude and 
$BP-RP$ color of stars within 250 parsec range from the sun to construct the Hertzsprung-Russell (HR) diagram \citep{1909HR_diagram, 1910HR_diagram, Russell_1914}. 
To control the signal-to-noise (S/N) ratio of data, we adopted the following criteria for our query:
\begin{description}
    \item \texttt{parallax\_over\_error > 10}
    \item \texttt{phot\_g\_mean\_flux\_over\_error > 50}
    \item \texttt{phot\_bp\_mean\_flux\_over\_error > 20}
    \item \texttt{phot\_rp\_mean\_flux\_over\_error > 20}
    \item \texttt{phot\_bp\_rp\_excess\_factor < 2}
    \item \texttt{ruwe < 1.2}
\end{description}
These criteria yield approximately two million {\it Gaia} sources for the color-magnitude diagram.

\subsection{GNIRS Near-Infrared Spectroscopy}

We obtained the near-infrared (NIR) spectroscopy of CSS1603+19 using the Gemini Near-Infrared Spectrograph 
\citep[GNIRS;][]{GNIRS_Design, 2006GNIRS_Spec_performance} instrument at the Gemini North Observatory on Mauna Kea, 
Hawaii (program ID GN-2019A-Q-322, PI: Hwang). We used the cross-dispersed mode with a 0.3-arcsecond slit, the 32 l mm$^{-1}$ grating, 
and the short blue camera (15$"$ pixel$^{-1}$) 
for the observation.  The setting ensured the S/N ratio is above 3, and the spectral resolving power is above 1600 to resolve 
emission lines and their radial velocity variations. On the nights of 2019-02-13, 2019-03-09, and 2019-03-17, we obtained four, nineteen, and fourteen 300-second exposures
respectively to cover the full orbital period of CSS1603+19. Our NIR spectrum covers wavelength range between 8500\,\AA\ and 26000\,\AA, giving full coverage over the $J$, $H$, and $K$ bands. 

We reduced the data using \texttt{PypeIt} v.1.8 and v.1.10 \citep{Prochaska2020Pypeit}. \texttt{PypeIt} uses A-B image differencing as the first step of sky subtraction, then improves 
sky subtraction further in individual 2D A-B difference spectra using sky apertures. It uses sky lines for wavelength calibration. We flux-calibrate using bright A0 stars observed right 
before and right after science exposures. The telluric absorption correction is computed from the spectrum itself assuming an underlying polynomial continuum. As a final step after all 
\texttt{PypeIt} reductions are done, we correct the wavelength calibration and the timestamps of the exposures to the heliocentric frame. 

\subsection{MDM Optical Spectroscopy}
\label{sec:MDM-spec}
We obtained optical spectra at the Michigan-Dartmouth-MIT (MDM) Observatory on Kitt Peak, Arizona \citep{Thorstensen_2017cv_period_note}. The 
observation used the ``Nellie" detector covering wavelength range between 4460\AA\; and 7770\AA. Bright O and B-stars were used to correct for 
telluric absorption. Comparison lamps and night sky line [OI]$\lambda$5577\AA\ were used to track night spectrograph flexure \citep{Thorstensen_2012cvperiods}.
A total of 32 600-second exposures were taken over multiple days in March and June 2017 in a high state. The raw data were reduced using IRAF routines. One-dimensional spectra were extracted using optimal-extraction algorithm of \citet{1986PASP_Horne_Spec_ex_algo}.

\subsection{APO Optical Spectroscopy}
We acquired more optical spectroscopy from the Apache Point Observatory (APO) at Sunspot, New Mexico. The observation was conducted on August 15, 2019 (PI: Hwang). The weather conditions were clear. We observed the source with Dual Imaging Spectrograph, where the red camera covers the wavelength from 6000\AA\ to 7000\AA. A 1.5-arcsec slit and the R1200 grating were used, with a resolving power $\sim1000$. The blue camera suffered from light contamination and did not provide scientifically useful data. We observed the target for $\sim$1.5 hours (covering about one full orbital period of CSS1603+19), with three 15-minute exposures and five 10-minute exposures bracketed by wavelength calibration. The data are reduced by the standard IRAF pipeline.

We were hoping to catch the object in a low state and detect the photosphere of the white dwarf in the APO observations, but caught it in an active (high) state, and we do not see any white dwarf photospheric features. We detect H$\alpha$ line and He I emission lines in the APO spectrum, but the following analysis focuses on the MDM spectra (Sec.~\ref{sec:MDM-spec}) for their larger wavelength coverage and a larger number of exposures. The overall behavior (e.g. radial velocity variations of H$\alpha$) of the APO spectra is consistent with that seen in the MDM spectra. 

\section{Measurements}
\label{sec:measurements}

This section describes the photometric and spectroscopic measurements for CSS1603+19. In Section \ref{sec:wise-photo}, we derive the object's orbital period from the mid-infrared light curves using WISE data. 
In Section \ref{sec:optical_photometry} we analyze optical photometry from ZTF and {\it Gaia}. In Sections \ref{sec:nir_cont} and \ref{sec:nir_spec} we present the continuum emission and strong line emission analysis for GNIRS near-infrared data, and in Section \ref{sec:opt_spec} we perform similar analysis but in the optical domain. All phase-resolved plots have the phase aligned with the WISE zero-phase point.

\begin{figure}
    \centering
    \includegraphics[width=\linewidth]{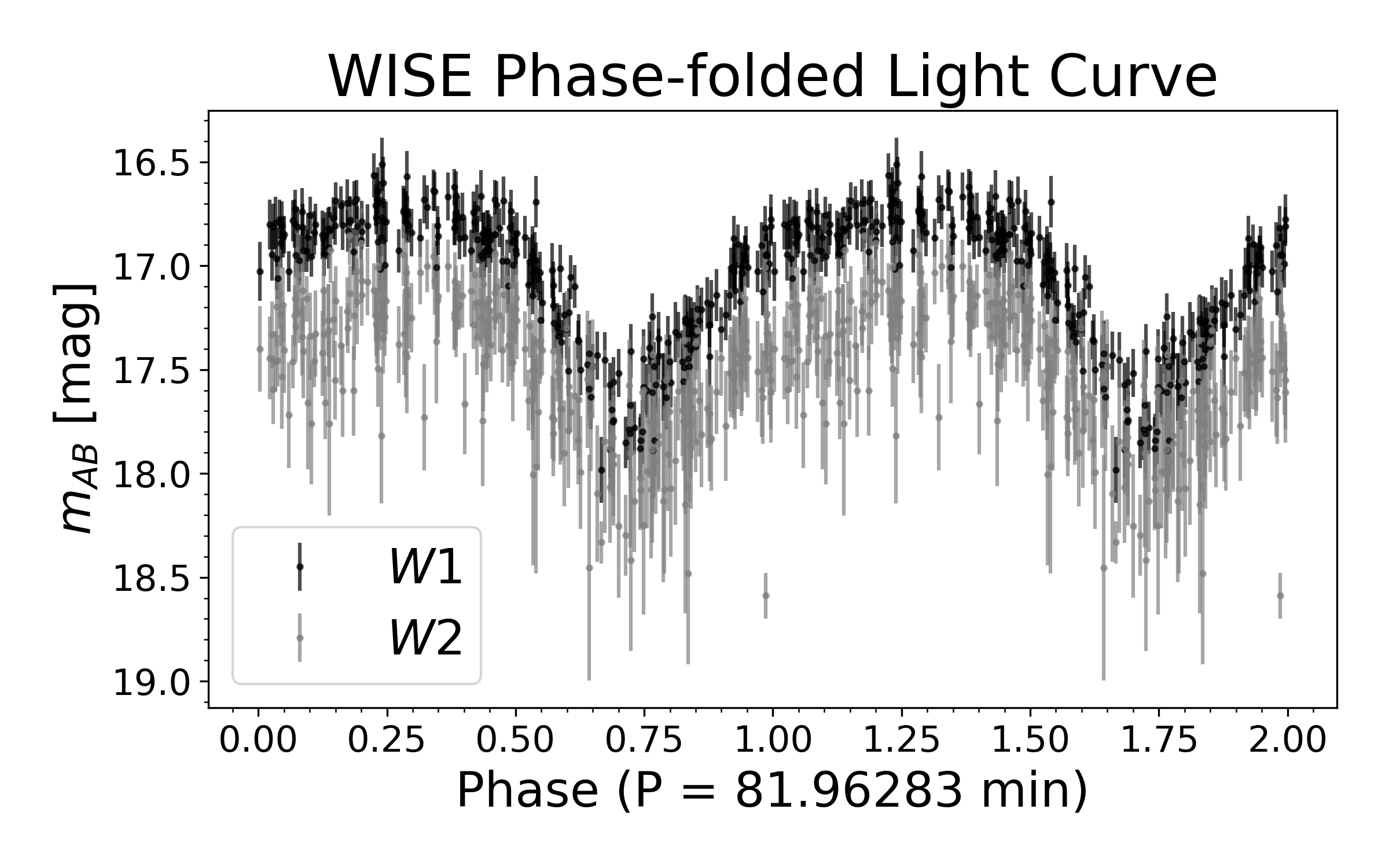}
    \caption{WISE light curve phase-folded with a period of 81.96283 min. Long-term variations are removed.}
    \label{fig:wise_light_curve}
\end{figure}

\subsection{Mid-Infrared Photometry}
\label{sec:wise-photo}

AllWISE and NEOWISE observations span $\sim 11$ years. The observed light curve reveals variability on a range of timescales. In particular, the light curve exhibits 
a long-term variation, with three high phases captured in 11 years where high phases are approximately one magnitude brighter than the low phase as shown in Figure \ref{fig:optical_raw} (right panel).

We search for periodicity after removing the long-term variability. Specifically, we group WISE photometry into 30-day groups. For each data point, we subtract the mean magnitude 
of its corresponding group and add the mean magnitude of all data points, thus removing the long-term variation from the WISE photometry.
Then, we use the Lomb-Scargle periodogram as implemented in \texttt{astropy} \citep{2018astropy} to search for the periodicity in the mean-calibrated WISE photometric data \citep{Lomb1976, Scargle1982}.
A prominent peak is detected, allowing us to derive the orbital period of CSS1603+19 of  $P_{\rm orb} = 0.05691863(25)$ day, 
in agreement with the period measurement $P = 0.05693(2)$\,day from \citet{Thorstensen_2017cv_period_note}, from the fits to the H$\alpha$ emission-line velocity.
We see no evidence of any change in the period over the 11-year timescale of WISE observations.

We obtain the phase-folded light curve of CSS1603+19 shown in Figure \ref{fig:wise_light_curve} by using $P_{\rm orb} = 0.05691863$ day. The v-shaped dimming feature has a minimum at orbital phase $0.701\pm 0.012$. In all subsequent figures we phase-fold light curves with the same period and phase zero point as the WISE observations so that the phases are directly comparable. 
By fitting the shape of the light curve with parametric functions or splines, we find that the object spends $45-50\%$ of the orbital period at flux values below the median flux. 
The peak-to-peak amplitude of the light curve is $1.1$ mag in W1 and $1.0$ mag in W2, but given the quality of the data this difference in depth is not statistically significant. We roughly split the WISE light curve into `high' and `low' states by considering separately the epoch fluxes above and below the median and we find that the eclipse depth and shape do not depend on whether the object is in the `high' or `low' state.  

\subsection{Optical Photometry}
\label{sec:optical_photometry}

Figure \ref{fig:optical_raw} shows the {\it Gaia} color-magnitude diagram. The {\it Gaia} sample consists of 2.2 million sources restricted to distances $< 250$ pc from the Sun.
The gold star marks the DR3 mean of the measurements for CSS1603+19 at $M_G = 11.855$ mag and $BP-RP = 0.356$ mag. The DR3 measurements are based on {\it Gaia} photometric measurements between Sept 2014 and Apr 2017.
This period captures an active high phase of CSS1603+19 starting June 2016. The active phase causes the epoch photometry of CSS1603+19 to be brighter and redder than the white dwarf sequence, as 
shown by the red dots in Figure \ref{fig:optical_raw}. The absolute $G$-band magnitude of CSS1603+19 reaches a minimum at around 12 mag, and when CSS1603+19 is in this low state no optical variability is detected. 

ZTF observatory provides $\sim 3$ years of optical photometry between 2018 and 2022. The observation captured two high states in Jan 2019 and Aug 
2019. The high states led to $\sim 2.5$ mag variability in the ZTF's $r$ and $i$ bands and $\sim 2$ mag variability in the $g$ band as shown in Figure \ref{fig:optical_raw} (right panel). 
We construct the phase-folded light curve using the ZTF photometric measurements in the low state. In contrast with mid-infrared measurements,
the ZTF photometric light curve shows no variability at the orbital period. 

\begin{figure*}
    \centering
    \includegraphics[width=\linewidth]{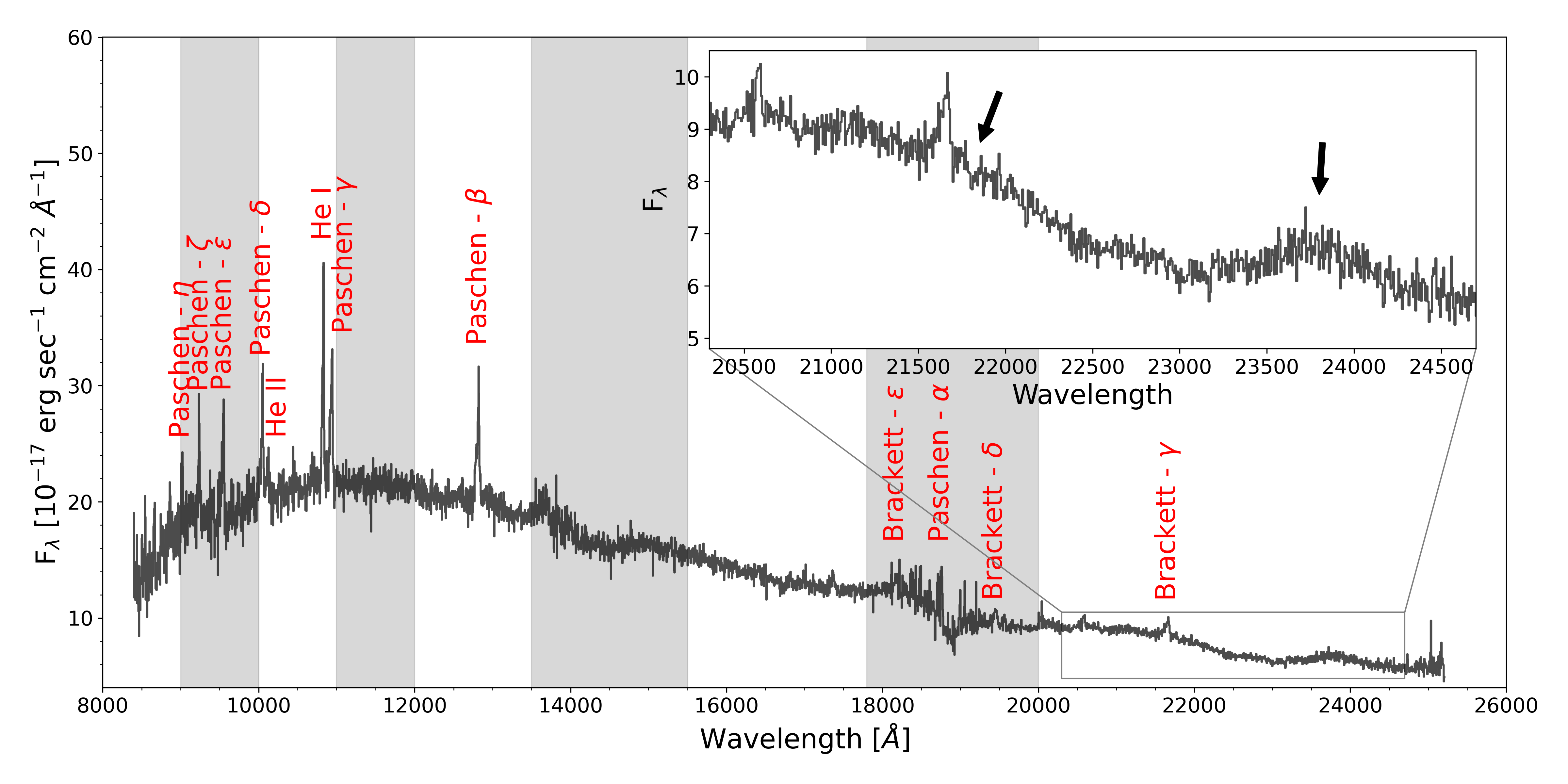}
    \caption{A typical GNIRS near-infrared exposure of CSS1603+19. The exposure was obtained on 2019-03-09. Hydrogen
    Paschen series emission lines are strong. Brackett series at high order are also visible within the wavelength range. He II at
    $\lambda$10094 and He I at $\lambda$10830 are also evident. Broad peaks of cyclotron emission are visible at long wavelengths (as indicated by the arrows and shown in the inset). Shaded regions are strongly affected by telluric absorption that may contain telluric residual after reduction.} 
    \label{fig:nir_spec}
\end{figure*}

\subsection{Near-Infrared Spectroscopy -- Continuum Features}
\label{sec:nir_cont}

A typical GNIRS spectroscopic exposure is shown in Figure \ref{fig:nir_spec}. We observe a time-varying continuum with strong emission lines and additional more subtle variations.

We fit a low-degree polynomial to the 
near-infrared continuum, masking the emission lines and overly noisy data points. We then calculate $JHK$-band magnitudes from the fits by integrating the model
spectral flux density within each band. We then phase-fold these synthetic continuum light curves using the WISE-derived period $P_{\rm orb}$ and its phase zero point, so that 
the phases in all the phase-folded light curves are directly comparable. The near-infrared bands show higher variability compared to the mid-infrared range:  
$J$ and $H$-band light curves both show peak-to-peak amplitude up to $\sim3$ mag.
The least variable $K$ band has peak-to-peak amplitude $\sim 2.5$ mag.

The continuum color is not constant. By computing the difference between the synthetic continuum fluxes, we find that the continuum becomes 
$\sim 0.4$ mag redder before the flux minima of the $JHK$ bands.

\begin{figure}
    \centering
    \includegraphics[width=\linewidth]{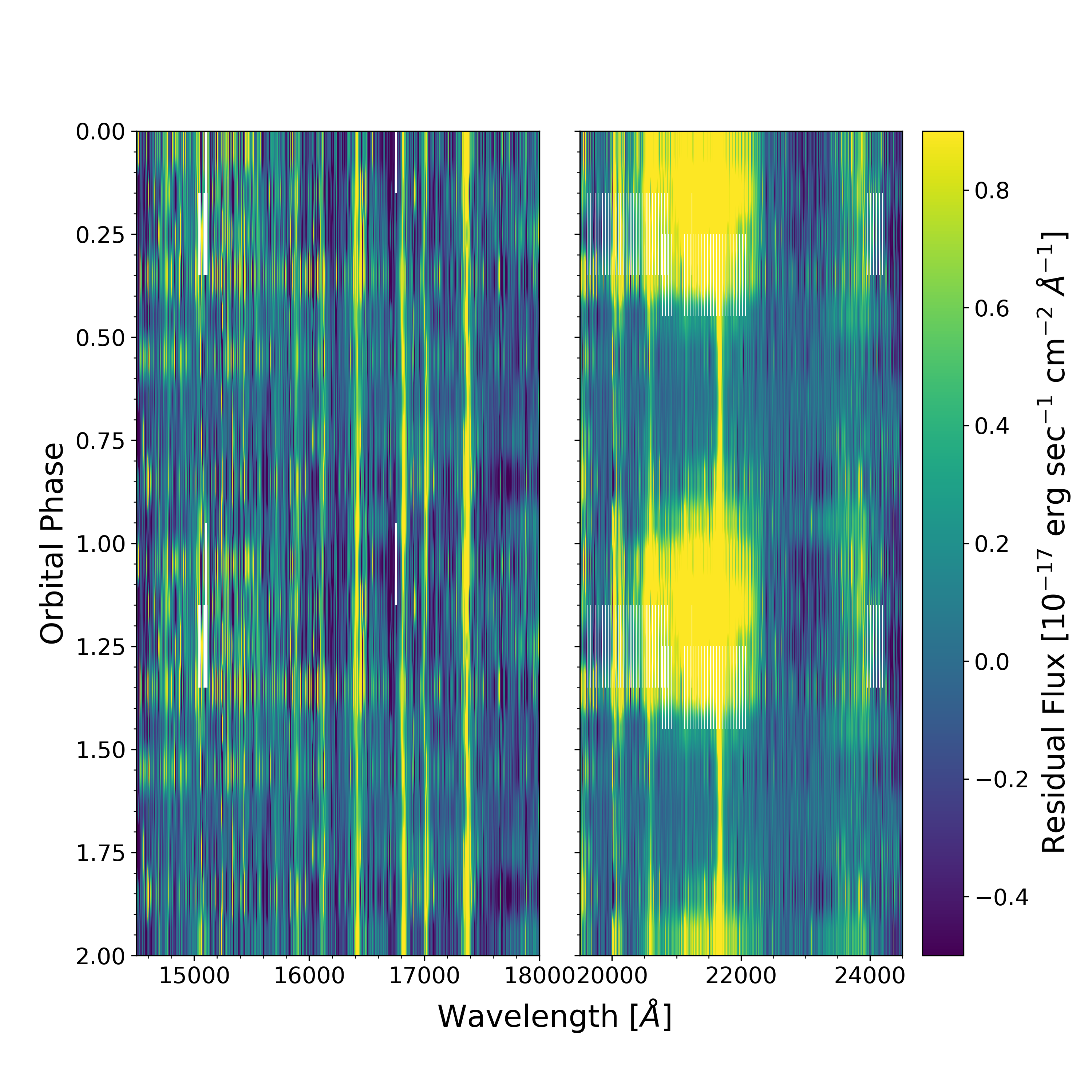}
    \caption{Phase-resolved spectroscopy which reveals cyclotron features in near infrared. Continuum is subtracted. 
    Clear near-infrared broad peaks appear during the bright state of the WISE light curve centered around 21400\AA\ and 23900\AA.
    }
    \label{fig:cyclotron}
\end{figure}

For some near-infrared spectroscopic exposures, we observe spectral wiggles between $\sim 20,500$\AA\ and $\sim 24,500$\AA\ as shown in Figure \ref{fig:nir_spec}.
These spectral features are likely cyclotron emission features originating in the strong magnetic field near the white dwarf surface. Subtracting 
the continuum to reveal these features, we find that they vary in intensity throughout the orbital period (Figure \ref{fig:cyclotron}), which is common 
for polars \citep{Schwope_1990_V834, Schwope_1993_MRSerp}. The NIR cyclotron feature is at its maximum intensity at the orbital phase $\varphi\simeq 0.13\pm 0.15$ 
and at its minimum intensity at $\varphi \simeq  0.63\pm 0.15$ -- i.e., approximately in phase with the WISE photometric intensity variation, which has 
a minimum at $\varphi=0.701\pm 0.012$. 

We recognize that if the donor is a brown dwarf, then its atmosphere may lead to broad spectral features in the longer wavelength domain in the near infrared spectroscopy due to the forest of molecular opacity features \citep{Ciardi1998}. In comparing the continuum-subtracted spectrum of CSS1603+19 with brown dwarf templates from \citet{2008Peterson_Brown_dwarf_atmosphere}, we find that the trough between the broad peak features in CSS1603+19 is inconsistent with the wavelength centroid of the dominant H$_2$O absorption in brown dwarfs. Furthermore, neither the overall flux of the features, nor their orbital evolution are consistent with a brown dwarf origin (Section \ref{sec:sed}). Thus we conclude that the atmosphere of the donor is unlikely to be responsible for these features.

In cgs units, the central wavelength of the $n$-th cyclotron harmonic is 
\begin{equation}
    \lambda_n=\frac{2\pi mc^2}{eBn},
    \label{eq:lam_cyclo}
\end{equation}
where $m$ and $e$ are electron's mass and charge and $B$ is the magnetic field. One feature is centered at $\sim 21,400$\AA\ and 
another at $\sim 23,900$\AA. If these features are due to two consecutive harmonics of the cyclotron emission, then this gives us two measurements for two unknowns ($n$ and $B$). Therefore, this relation allows us to estimate $n\simeq 8-10$ and $B\simeq 5-6$ MG. We do not detect any other cyclotron harmonics at shorter wavelength; our ability to do so may be affected by telluric corrections. We discuss the cyclotron scenario in more detail in Sec.~\ref{sec:discussion-ir}.

\subsection{Near-Infrared Spectroscopy: Emission Lines}
\label{sec:nir_spec}

\begin{figure*}
    \centering
    \includegraphics[width=\linewidth]{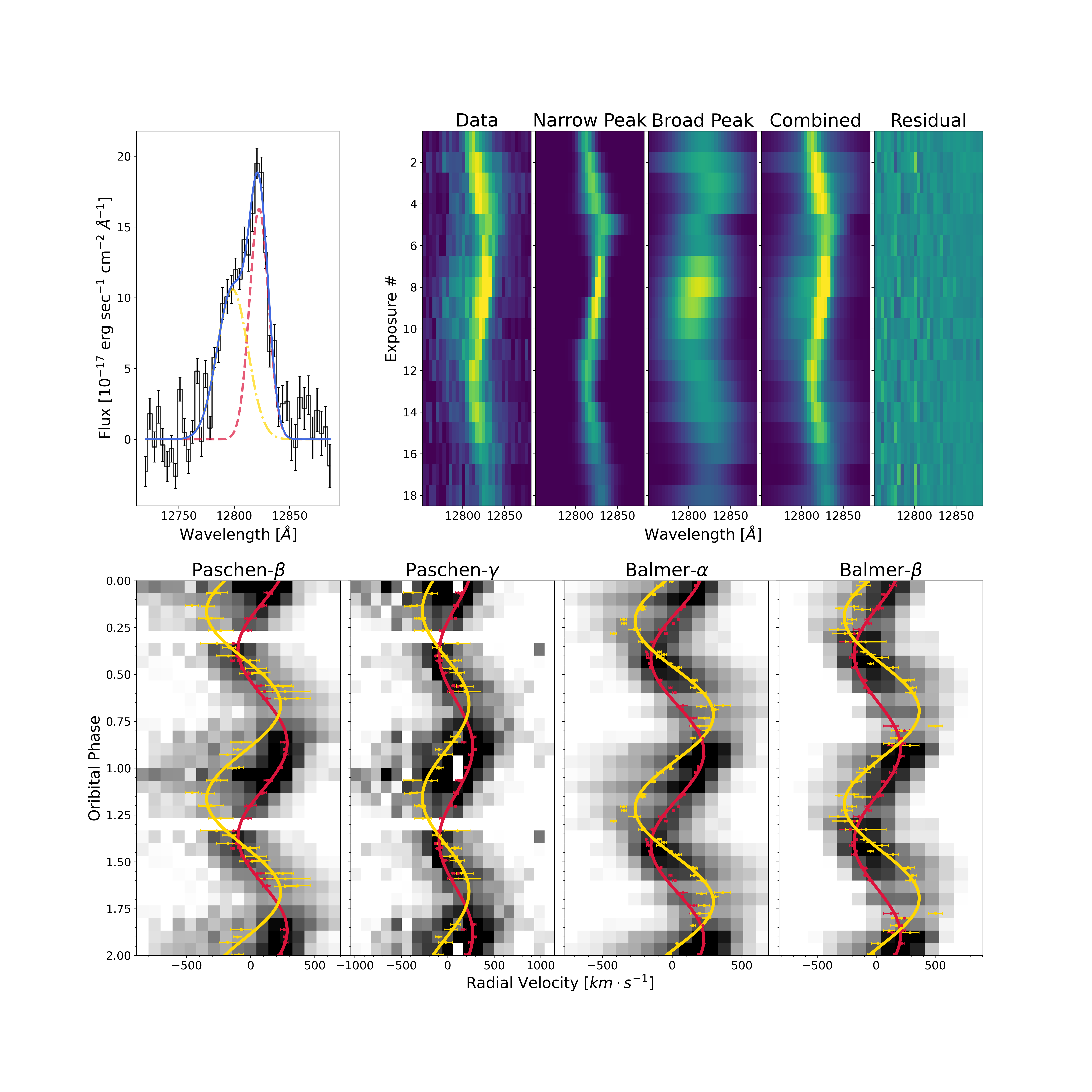}
    \caption{Top left: double-Gaussian fit to continuum-subtracted Paschen $\beta$ emission line in one exposure, with data in black and overall fit in blue 
    line. The fit for the broad peak is shown with the yellow dot-dashed curve, and the narrow component with the red dashed curve. 
    Top right: Paschen $\beta$ orbital evolution and fits on 2019-03-09. 
    From left to right, the panels are raw data, narrow peak fit, broad peak fit, double peak fit, and residual significance (maxed at S/N=3), with yellow corresponding to bright emission and blue corresponding to faint emission. The double peak emission is most obvious for exposures 8-11 where the broad component presents on the blue (short wavelength) side of 
    the narrow peak, and for exposure 1-4 where the broad component is on the side of the narrow peak.
    Bottom row, from left to right: phase-resolved spectra of Paschen $\beta$, Paschen $\gamma$, Balmer $\alpha$, and Balmer $\beta$ respectively, with black corresponding to bright emission and white corresponding to faint emission. One NIR exposure is removed due to cosmic ray contamination. Continuum is subtracted. Two emission components are visible. The narrow component is traced by the red sinusoidal
    curve; the yellow curve traces the broad component. The peak wavelengths of the two peaks obtained from individual fits are marked by points with error bars. The broad component has a larger velocity semi-amplitude than the narrow component and precedes the narrow component by approximately 0.2 of the orbital phase.}
    \label{fig:double_peak_lines}
\end{figure*}

All identified strong lines in Figure \ref{fig:nir_spec} show similar orbital variations of their radial velocity profile. After excluding the lines whose fitting quality is affected by the continuum 
emission and other emission mechanisms, we model five emission lines (Paschen $\beta$, $\gamma$, $\delta$, $\epsilon$, and Brackett $\gamma$) to derive 
their radial velocities and other kinematic measures. 

We first fit emission lines using a Gaussian profile. The radial velocities derived from the Gaussian centroid of each peak are 
noise-weighted for each exposure. To model the orbital evolution of the emission lines, we then perform non-linear least squares fit for the noise-weighted radial velocity using a sinusoidal model in the phase domain:
\begin{align}
    \label{eq:sin_model}
    V(\varphi\;|\;\gamma, K, \phi) = \gamma + K\sin(2\pi(\varphi - \phi))
\end{align}
where $K$ is the velocity semi-amplitude, $\gamma$ is the systematic velocity and $\phi$ is the overall phase shift.
We minimize the number of free parameters for the model to maximize reliability of the radial velocity measurement, for example by fixing the orbital 
period to that derived from WISE photometry. The model fitting yields semi-amplitude $K = 190.1\pm 7.6$ km s$^{-1}$, 
systemic velocity $\gamma = 83.3\pm 5.3$ km s$^{-1}$, and phase of the velocity node $\phi = 0.5046\pm 0.0059$.

A closer look at the emission lines reveals interesting substructure. In Figure \ref{fig:double_peak_lines}, we show the phase-resolved spectra of Paschen 
emission lines where we split the orbital period into 15 bins. Aside from the stronger component traced by the red line,
a fainter component is also visible. Accurate characterization of the faint 
component needs a high S/N ratio and minimal effects from 
nearby emission peaks and continuum noise. Therefore, while we detect similar double peaked emission from both hydrogen and helium lines, 
we only analyze the double-peaked structure of the strongest lines -- Paschen $\beta$ and Paschen $\gamma$ -- in the near-infrared domain. 

We use a double-Gaussian model to fit the emission lines in all exposures via least squares optimization, fixing the ratio of the peak amplitudes throughout the orbit. The time evolution of each component is modeled using the sinusoid from Equation \ref{eq:sin_model}. 
Velocity dispersion for both components are free to vary in the time-dependent fit. The double-Gaussian fitting of the emission lines throughout the orbit demonstrates that the velocity dispersion of the narrow peak is consistently smaller ($\sim 150$ km s$^{-1}$) than that of the broad peak ($\sim 350-600$ km s$^{-1}$). At the spectral resolution $R\gtrsim 1600$, the emission lines are well spectrally resolved, so these values reflect the spread of physical velocities of gas responsible for the emission. Figure \ref{fig:double_peak_lines} shows a typical fit for one exposure for the Paschen $\beta$ line and the overall fit quality for 18 exposures on 2019-03-09. One exposure is removed due to potential cosmic ray contamination. The results are summarized in Table \ref{tab:double_gauss_result}.

The broad component and the narrow component are systematically offset from one another by about 0.2 in orbital phase. To verify the reliability of this measurement, we fit the entire ensemble of emission lines in the velocity phase space using the model in Equation \ref{eq:2d_line_fit}.
The Equation describes the double peak combination in given phase bin $\varphi$ is as follows:
\begin{equation}
    \begin{aligned}
        \label{eq:2d_line_fit}
        I(v | \varphi) = &\sum_{i\in\{n, b\}} c_iI_\varphi
        \exp\left(-\frac{(v - V(\varphi\;|\;\gamma_i, K_i, \phi_i))^2}{2\sigma_i^2}\right)
    \end{aligned}    
\end{equation}
where $I_\varphi, \gamma_i, K_i, \phi_i$ are parameters need to be fitted. We use $N_{\rm bin}=15$ bins for constructing the phase-resolved spectroscopy, 
therefore we have $N_{\rm bin} + 6$ free parameters in total. Here $i\in \{n, b\}$ are the indices distinguishing the narrow and the broad Gaussian components, 
and $I_\varphi$ is the amplitude of the narrow peak the given phase bin $\varphi$. Radial velocity $V(\varphi\;|\;\gamma_i, K_i, \phi)$ has the same 
sinusoidal definition as in Equation \ref{eq:sin_model} to model the peak position across the entire orbital phase, but now we are fitting all spectra as 
an ensemble. $c_i$ are constants used for fixing the amplitude ratio between narrow and broad peaks. We use $c_n = 1$ and $c_b = 0.65$ for a reasonable fitting. 
$\sigma_i$ are the widths of the narrow and broad Gaussian components. 

We fix the Gaussian widths by using the median values we obtained using prior fitting of all spectra independently, which allows us to limit the number 
of free parameters without significantly harming the overall fitting quality. On the contrary, the peak amplitude shows significant variation over the 
orbital phase. Fixing the broad amplitude introduces large fitting residuals, defeating our goal to verify the systematic velocity measure, so we allow 
it to vary. The fits for Paschen $\beta$ and Paschen $\gamma$ confirm that the phase offsets between the narrow and broad components are $0.203(14)$ and 
$0.189(22)$ respectively -- i.e., close to, but not quite a quarter of the orbital period, with the phase difference of 0.05 from the quarter of a period 
now verified using ensemble fitting at high statistical significance. 

The net systematic radial velocities of the broad and narrow components appear to be different as well. To verify the difference of systematic velocity, 
we compare the fitting result with and without requiring $\gamma_n = \gamma_b$. For the two Paschen lines, both the Akaike information criterion  
and the Bayesian information criterion show significant favor to the model whose $\gamma_n \neq \gamma_b$. The broad peaks are systematically blue-shifted by $212\pm15$ km s$^{-1}$ for Paschen $\beta$
and $176\pm 29$ km s$^{-1}$ for Paschen $\gamma$ compared to the narrow peaks. There are differences between the ensemble fitting results and 
the results from fitting individual exposures summarized in Table \ref{tab:double_gauss_result}. 
While the ensemble fitting is better for demonstrating the statistical significance of the phase difference and systematic velocity difference between the narrow and broad components of the emission lines, our analysis below is based on the fitting of individual 
exposures in which fewer constraints are imposed over the fit.

\begin{table*}[t]
    \centering
    \caption{Double-Gaussian emission line fit result. The mask column indicates the orbital phase masked that is not considered during fitting. Component indicates
    the broad (B) and narrow (N) emission component of the emission line. $\sigma$ is the mean of velocity dispersion values obtained in individual fits.}
    \begin{tabular}{ccccccc}
    \hline
         Strong Line& Mask 
         & Comp.
         & $K$ (km s$^{-1})$
         & $\gamma$ (km s$^{-1})$
         & $\phi$
         & $\sigma$ (km s$^{-1}$)\\
    \hline
    \multirow{2}{*}{Paschen $\beta$} & \multirow{2}{*}{$0.65< \phi < 0.85$}  & N & $192.1\pm 7.5$ & $93.0\pm 5.3$ & $0.6172\pm0.0074$& $190\pm 69$\\
                  
                  & & B & $289\pm28$ & $-55\pm17$ & $0.414\pm 0.011$ & $620\pm 110$\\
    \hline
    \multirow{2}{*}{Paschen $\gamma$} & \multirow{2}{*}{$0.6 < \phi < 0.8$} & N & $181.9\pm10.0$ & $87.6\pm7.7$ & $0.638\pm0.011$& $173\pm 74$\\
                 & & B & $251\pm25$ & $-22\pm14$ & $0.408\pm0.011$& $600\pm140$\\
                 \hline
    \multirow{2}{*}{Balmer $\alpha$} & \multirow{2}{*}{$0.65< \phi < 0.75$}  & N & $196.3\pm 6.1$ & $9.4\pm 4.4$ & $0.6433\pm0.0053$ & $154\pm 36$\\
                  & & B & $318\pm21$ & $47\pm13$ & $0.4438\pm 0.0080$& $360\pm73$\\
    \hline
    \multirow{2}{*}{Balmer $\beta$} & \multirow{2}{*}{$0.85 < \phi < 0.9$} & N & $189.7\pm 6.4$ & $37.9\pm 5.3$ & $0.6640\pm0.0069$& $169\pm 38$\\
                 & & B & $281\pm17$ & $15\pm11$ & $0.4655\pm0.0070$ & $365\pm81$\\
    \hline
    \end{tabular}
    \label{tab:double_gauss_result}
\end{table*}

\subsection{Optical Spectroscopy}
\label{sec:opt_spec}

\begin{figure*}
    \centering
    \includegraphics[width=\linewidth]{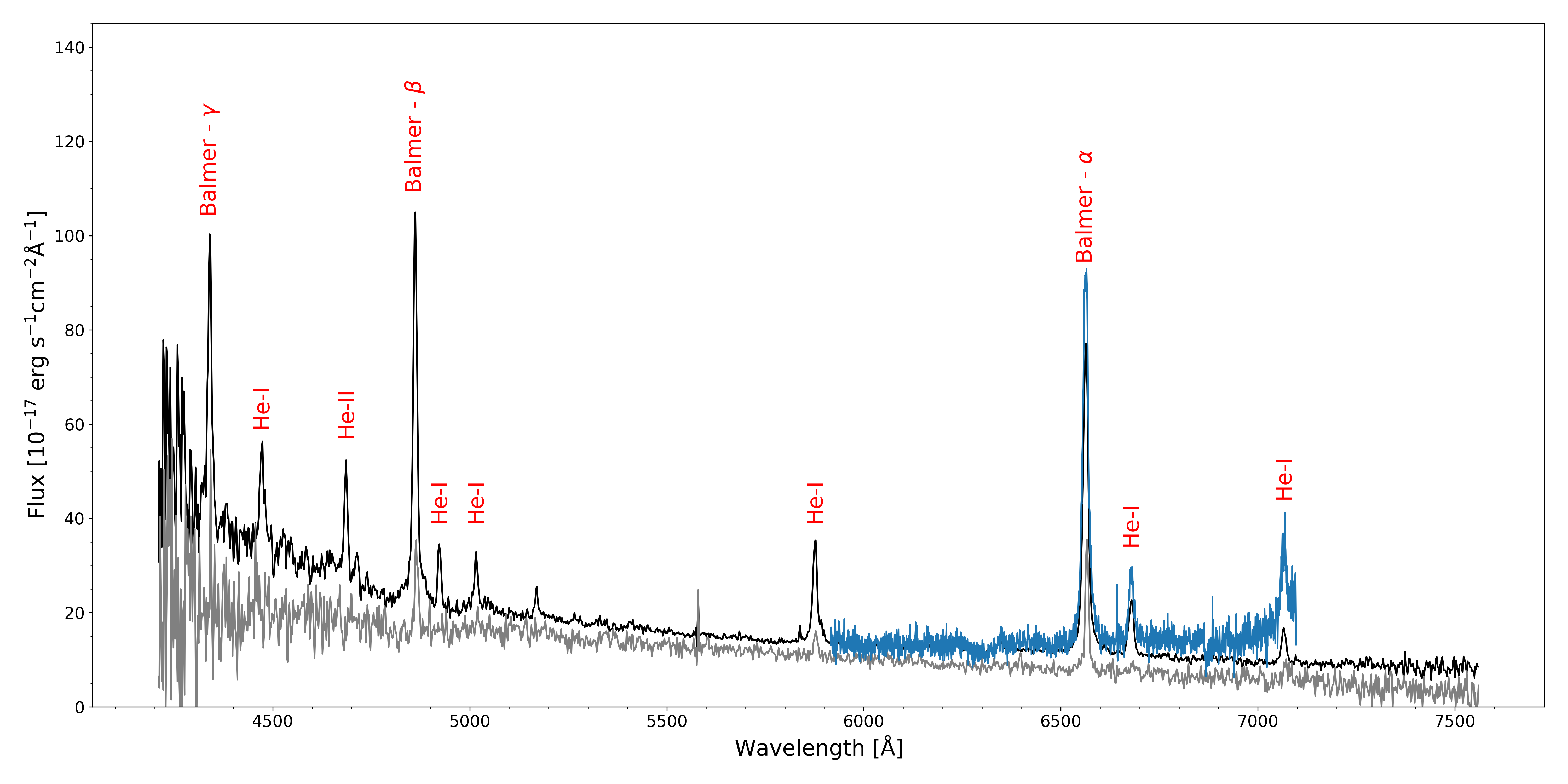}
    \caption{Mean optical spectra: black for the March 2017 MDM observation, gray for the June 2017 MDM observation and blue for the August 2019 APO observation. Balmer series, He I, and He II $\lambda$4686\AA\ are clear, but the He II line becomes undetectable during the June 2017 observation. The continuum emission is also fainter in June 2017.}
    \label{fig:optical_spec}
\end{figure*}

The optical spectrum shows clear continuum emission and strong emission lines as exhibited in Figure \ref{fig:optical_spec}. Using the same approach used for near-infrared analysis, we fit a low-degree polynomial to approximate the continuum emission while masking the emission lines. The continuum emission is more stable over the optical range compared to its near-infrared counterpart. The continuum flux remained approximately constant over the orbital period. No significant pattern is recognized in the synthetic $V$-band and $R$-band light curves.  

In addition to the strong Balmer series of hydrogen, He I lines and He II $\lambda$4686\AA\ are evident in the optical range during the March 2017 observation. The He II line becomes undetectable in June 2017 when CSS1603+19 enters a low phase. Even in the faintest optical spectrum we have obtained, the white dwarf photosphere is not discernible, unlike the case of some CVs with extremely low accretion rates \citep{Gaensicke2009}. 

The two-component emission-line structure is also observed in the optical spectrum. We first fit for the overall orbital evolution of the radial velocity without deblending the two components using the model described by Equation \ref{eq:sin_model}. We include Balmer $\alpha, \beta$, He I $\lambda$5016\AA, 
and He I $\lambda$6678\AA. We find semi-amplitude $K = 197.0\pm 5.2$ km s$^{-1}$, 
systemic shift $\gamma = 35.4\pm 3.4$ km s$^{-1}$, and phase of the velocity node $\phi = 0.4549\pm 0.0041$. 
The semi-amplitude is slightly larger than the measurement using near-infrared data, but the difference is within the uncertainty of the measurement.

We then deblend the broad and the narrow components of emission lines using the same method as in Section \ref{sec:nir_spec}, fitting Balmer $\alpha$ and Balmer $\beta$. The fitting result is summarized in Table \ref{tab:double_gauss_result} and visualized in  Figure \ref{fig:double_peak_lines}. Similar to the near-infrared case, the broad emission peak precedes the narrow emission peak by 0.2 orbital phase up to a high accuracy. In contrast with Paschen $\beta$ and Paschen $\gamma$, the broad components of Balmer lines do not have a blue systematic shift. By fitting the ensemble of spectra using Equation \ref{eq:2d_line_fit},
we do not find any evidence for a systematic offset between broad and narrow components of Balmer $\alpha$. The Balmer $\beta$ line has its broad component systematically redshifted by $37\pm 9.0$ km s$^{-1}$ compared to the narrow component. This is in contrast to the systematic blueshift we see for the Paschen sequence. 

\section{Analysis and Discussion}
\label{sec:analysis}

This section focuses on the physical interpretation and analysis of measurements from Section \ref{sec:measurements}.

\subsection{Spectral Energy Distribution (SED) Analysis}
\label{sec:sed}

\begin{figure}
    \centering
    \includegraphics[width=\linewidth]{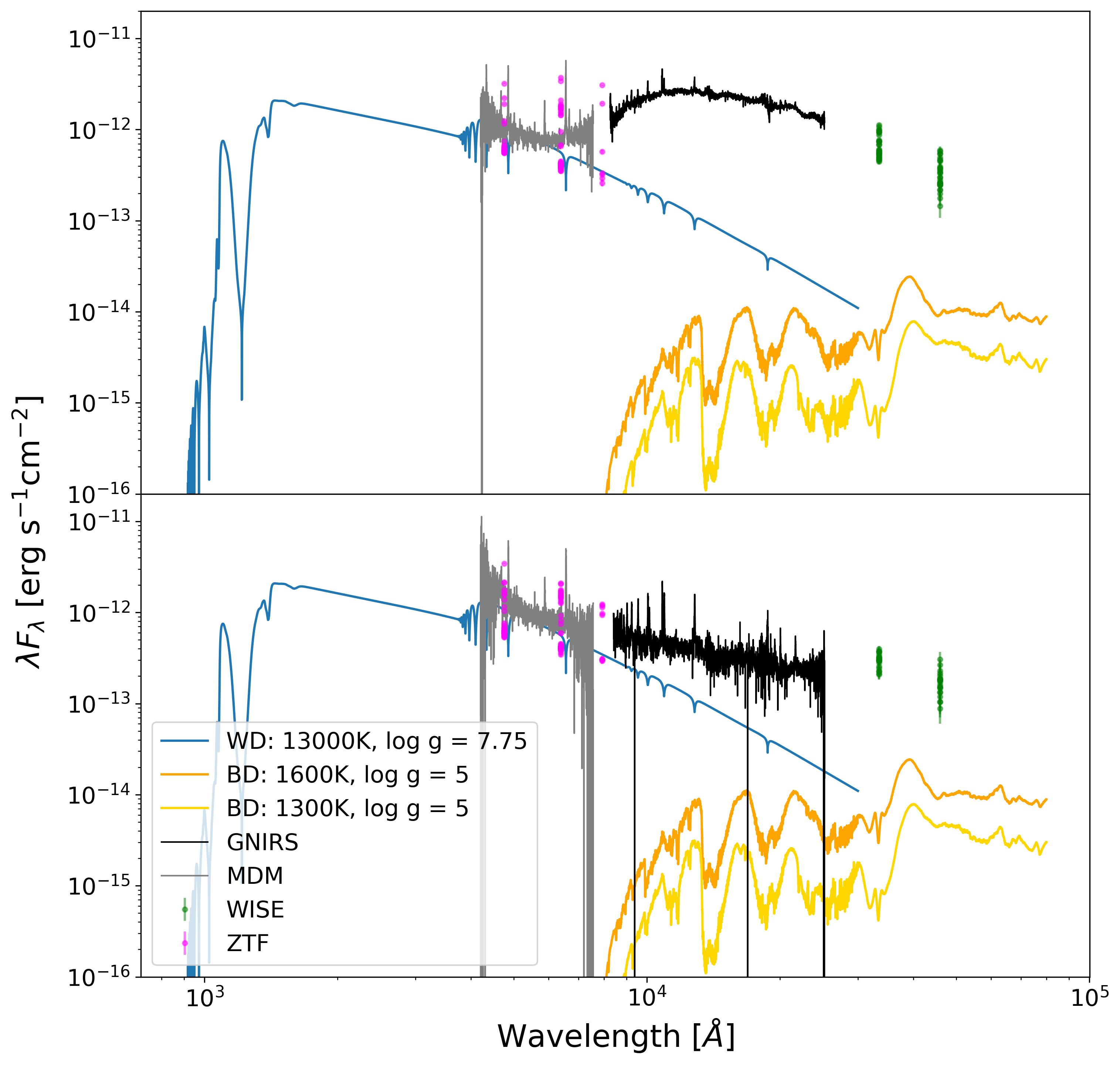}
    \caption{SED of CSS1603+19 in the bright phase (orbital phase bin $0.15 < \varphi < 0.25$, top) and in the faint phase ($0.65 < \varphi < 0.75$, bottom). The colored dots are observations labeled as by legends. The blue curve shows a 13,000 K white dwarf model
    with surface gravity $\log g = 7.75$ from \citet{2010Koester_wd_atm_model} and \citet{2009Tremblay_wd_atm_model} as our white dwarf estimate. 
    The orange and gold curve showed the 0.1 R$_\odot$ brown dwarf emission model at 1600K and 1300K respectively using templates from \citet{2021Marley_BD_template}.}
    \label{fig:ir_sed_BB_fit}
\end{figure}

Using optical and infrared data from WISE and ZTF and our own spectroscopic measurements from MDM and Gemini, we construct the spectral energy distribution (SED) of CSS1603+19 shown in Figure \ref{fig:ir_sed_BB_fit}. We select available observations that fall into a relatively bright orbital phase ($0.15 < \varphi < 0.25$) and into a relatively faint orbital phase ($0.65 < \varphi < 0.75$). We observed significant variability in the infrared. On the contrary, the variability in the optical is uncorrelated with the orbital phase, and this is also apparent from Figure \ref{fig:ir_sed_BB_fit}. The SED shows a clear infrared excess and a local minimum at around $\lambda \simeq 8000$\AA. 

We first attempt to model the SED of CSS1603+19 with a sum of spectra of a white dwarf and a brown dwarf. We select a $13000K$, $\log g = 7.75$ white dwarf template from \citet{2010Koester_wd_atm_model} and \citet{2009Tremblay_wd_atm_model} to model the white dwarf emission. In some cases, the brown dwarf atmospheric absorption bands may lead to spectral features between 2 \micron\ and 2.4 \micron\ \citep{Ciardi1998} qualitatively similar to those seen in our data. Thus, we use brown dwarf atmosphere templates from \citet{2021Marley_BD_template} spanning a temperature range of 800K to 2400K with surface gravity $\log g = 5$ to check whether our spectral wiggles match the brown dwarf absorption features. The brown dwarf H$_2$O absorption leads to a broad peak between 2.1 \micron\ and 2.3 \micron\ whose exact centroid is dependent on the brown dwarf temperature. To match the peak at 2.14 \micron\ observed in our target, the brown dwarf temperature needs to be between the 1300 and 1600\,K range. 

Upon determining the temperature, we normalize the spectrum using the brown dwarf radius $0.1R_\odot$, leading to the results shown in figure \ref{fig:ir_sed_BB_fit}. By comparing the model spectral peak and the observed spectral peak, we find that the flux of the peak in the 1600\,K 0.1 $R_\odot$ model is too faint by a factor of 10 compared to the observed value, and the flux in the 1300K model is too weak by a factor of 30. Scaling up the brown dwarf to match the model flux within the peak to the observed one yields a minimum brown dwarf radius $\sim$0.35 $R_\odot$, which is nonphysical. Apart from the peak at 2.14 \micron, the brown dwarf model also predicts two other peaks near 1.65 \micron\ and 1.3 \micron, which are undetected in our NIR spectroscopy, while the observed peak at 2.39 \micron\ peak in our spectroscopy remains unaccounted for by the brown dwarf template. 

The overall luminosity of our model brown dwarfs is too faint by about two orders of magnitude compared to the observed infrared continuum in the bright phase. In CVs, the donor star is significantly larger than the white dwarf. Thus, the large periodic variability of the 2.14 \micron\ feature and the NIR continuum are also unexplained in a model where the NIR flux is dominated by the brown dwarf, as there is no occulting body with the size comparable to the size of the brown dwarf. 

In some CVs, there is a circumbinary disk that can produce infrared excess. 
The size of the circumbinary disk may help explain the strong infrared component in the spectral energy distribution which cannot be accounted for by the donor star (e.g. \citealt{2004Dubus_mid_IR_excess}). However, the geometry of
the circumbinary disk is not compatible with the eclipsing-like infrared lightcurve and rapid variability. The infrared emission from a circumbinary disk can only be minimally 
blocked by the binary, and the geometry prevents the observed factor of two variability. Thus, the model cannot explain the large 1-mag infrared variability in CSS1603+19. Several CVs suspected of having circumbinary disks were later shown to have their infrared excess dominated by the donor or by cyclotron emission \citep{Harrison2013, 2004Dubus_mid_IR_excess}.

\subsection{Binary model}
\label{sec:binary-model}

\begin{figure}
    \centering
    \includegraphics[width=\linewidth]{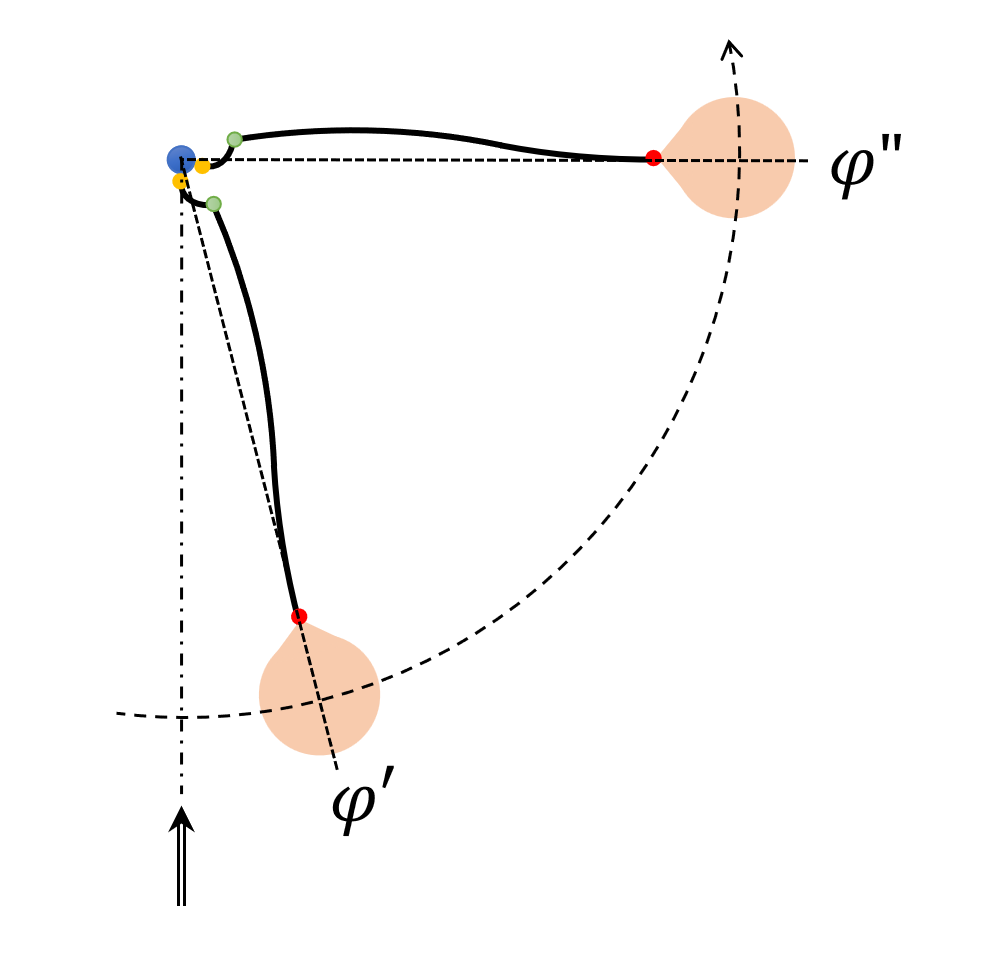}
    \caption{The geometry of CSS1603+19. The observer is fixed below the figure with the line of sight indicated by the arrow, while the system rotates counterclockwise. Gold and red dots represent the line-emitting regions producing the broad and narrow components, respectively. The same color scheme is used in Figure \ref{fig:double_peak_lines} to illustrate the radial velocity behavior of these lines. The green spot marks the location where the stream's geometry becomes shaped by the white dwarf magnetic field. $\varphi'=0.6874\pm 0.0089$ is the orbital phase where the broad component appears with the largest redshift. $\varphi'' = 0.8915\pm 0.0071$ is the orbital phase where the narrow component appears with the largest redshift. The proposed geometry of the curved accretion stream results in $\varphi'' - \varphi' = 0.204\pm 0.016$ (instead of the 0.25 if the stream were straight). While the white dwarf is also moving relative to the center of mass of the system, its mass is likely much larger than that of the donor, and therefore we illustrate it as nearly stationary. The orbital inclination is not incorporated in the figure.}
    \label{fig:cartoon_illustration}
\end{figure}

\begin{table}
    \centering
    \caption{Stellar Parameters}
    \begin{tabular}{l}
        \Xhline{3\arrayrulewidth}
        \multicolumn{1}{c}{\textbf{System Parameters}}\\
            $P = 81.96282\;\pm\;0.00036 \textrm{\;min}$\\
            $d = 235.3_{-6.4}^{+6.9} \textrm{\;pc}$\\
            $i \lesssim 74.6\pm 0.4 ^{\circ}$\\
            $q \lesssim 0.263 \pm 0.046$\\
            $a \gtrsim 0.425 \pm 0.024R_\odot$\\
        \multicolumn{1}{c}{\textbf{Stellar Mass Radius}}\\
            $M_1 \gtrsim 0.245\pm 0.032 M_\odot$\\
            $R_1 \lesssim 0.01241\pm 0.00023 R_\odot$\\
            $M_2 = 0.0644\pm 0.0074 M_\odot$\\
            $R_2 = 0.113\pm 0.013 R_\odot$\\
        \multicolumn{1}{c}{\textbf{Cyclotron Region}}\\
            $B = 5 - 6MG$\\
        \Xhline{3\arrayrulewidth}
    \end{tabular}
    \label{tab:star_params}
\end{table}

In this section, we constrain the stellar parameters of our binary system and model a range of observables we discussed above. The first striking feature that gives us a clue as to the geometry of the system is the radial velocity variations of the two peaks of the emission lines which are out of phase and have different amplitudes. Furthermore, the semi-amplitude of the broad component $K_{b}\simeq 300$ km s$^{-1}$ is high enough that the locus of the possible solutions in the space of white dwarf ($M_1$) and donor companion ($M_2$) masses suggests unrealistically high masses for the companion ($\ga 0.5 M_{\odot}$) which we would have seen in our data. This observation is reminiscent of that by \citet{Mason2019} for  eclipsing polar CRTS J0350+3232 with an orbital period of 2.4 hours. We therefore adapt the geometric model of \citet{Mason2019} for our system, albeit with some modifications as described below. 

Specifically, following \citet{Mason2019} we hypothesize that the broad component of the emission line originates from the accretion spot close to the surface of the white dwarf. Its large radial velocity amplitude is due to a net flow of matter within the accretion stream toward the white dwarf. Figure \ref{fig:cartoon_illustration} illustrates that $\varphi'$ is the orbital phase where the radial velocity of the broad component (gold spot in the figure) is at its maximal redshift. The narrow component (red dot in Figure \ref{fig:cartoon_illustration}) is placed nearer to the donor, although following the investigation of the resulting $M_1-M_2$ solutions we are inclined to place it at the Roche point rather than at the exact orbital location of the donor. 

If the stream connecting the donor and the white dwarfs was straight, then just by associating one end of it with the Roche point and the other end of it with the accretion spot we would expect an 0.25 phase offset between the two components. In practice, due to the Coriolis force, the stream bends in the same direction as the orbital motion, and the resulting trajectory was computed by \citet{1975Lubow_accretion_stream}. Associating one end of such stream with the Roche point and the other end with the accretion spot, one would expect an $>0.25$ phase offset between two components assuming the absence of an accretion disk. Instead the phase offset is $\varphi'' - \varphi' = 0.204\pm 0.016$. This suggests that the accreting stream lags the donor star by $\sim 18$ degrees (the same lag is seen in the polars described by \citealt{Mason2019} and \citealt{Tovmassian1997}). The bending of the stream required by these observations is possible for a polar where the accretion geometry close to the white dwarf is dominated by its magnetic field. Following the AM UMa model depicted in \citet{Bonnet1996}, we show in Figure \ref{fig:cartoon_illustration} that as the stream propagates from the donor to the white dwarf, it first curves in the direction predicted by \citet{1975Lubow_accretion_stream} and then twists due to the effects of the magnetic field to account for the observed phase differences in the emission lines. Finally, the $\sim 100$ km s$^{-1}$ systemic blueshift of the broad component relative to the narrow component seen in some of the lines (Pa $\alpha$, Pa $\beta$ as listed in Table \ref{tab:double_gauss_result}) may be due to partial obscuration of the accretion spot by the stream at phase $\varphi'$, when the emitting component is at its highest redshift. 

The possible cyclotron features in the near-infrared spectroscopy indicate the presence of strong magnetic fields from the system, likely associated with the white dwarf. The strong magnetic field near the white dwarf may interrupt the accretion disk, causing the accretion stream to flow
along the magnetic field lines. The presence of stable weak broad components to the emission lines that show regular orbital variations implies a stable accretion flow near the white dwarf. This indicates the white dwarf rotates synchronously with respect to the 
system, revealing the polar (AM Her star) nature of the system in contrast to an intermediate polar (DQ Her Type) in which the white dwarf 
rotates asynchronously \citep{1990Cropper_Polar_Review}. Abundant emission lines and the strong He II at $4686$\AA\ also 
mark the system as an AM Her type CV, although the system does not quite meet the HeII$4686$\AA/H$\beta>0.4$ criterion considered sufficient for classifying a system as a magnetic CV \citep{Silber1992, Mason2019}. 

Our first dynamical constraint comes from the size of the Roche lobe. The $81.96$-min orbital period of CSS1603+19 suggests that the system is below the CV period gap. As accretion signatures are clearly visible, the donor must be overflowing its Roche lobe whose radius is completely determined by the semi-major axis and the mass ratio of the system \citep{1983Eggleton_Roche_radius}. Fully convective stars -- such as the donor below the period gap -- are well described by $n = 1.5$ polytropes.
Thus, the donor radius can be constrained using the donor mass-radius relation and the 
Roche-lobe radius approximation:
\begin{align}
    \label{eq:eggleton_roche_lobe}
    \frac{R_2}{a} = \frac{0.5126q^{0.7388}}{0.6710q^{0.7349} + \ln(1 + q^{0.3983})},
\end{align}
in which $R_2$ is the donor radius, $a$ is the the semi-major axis, and $q=M_2/M_1$ is the mass ratio 
\citep{2009Sirotkin_Roche_approx, 2011Knigge_CV_evo}.
If we instead assume that the system is a period bouncer, then the period bouncer mass-radius relation from \citet{2011Knigge_CV_evo} yields a donor mass of 0.1 $M_{\odot}$, somewhat outside of the range of applicability of the mass-radius relation for bouncers ($M<0.069 M_{\odot}$), and we therefore do not obtain a self-consistent solution. We thus conclude that our target has likely not yet entered the period-bouncer stage of its evolution; however, our derived mass is close to the bouncer range and therefore this conclusion is somewhat uncertain. 

\begin{figure*}
    \centering
    \includegraphics[width=\linewidth]{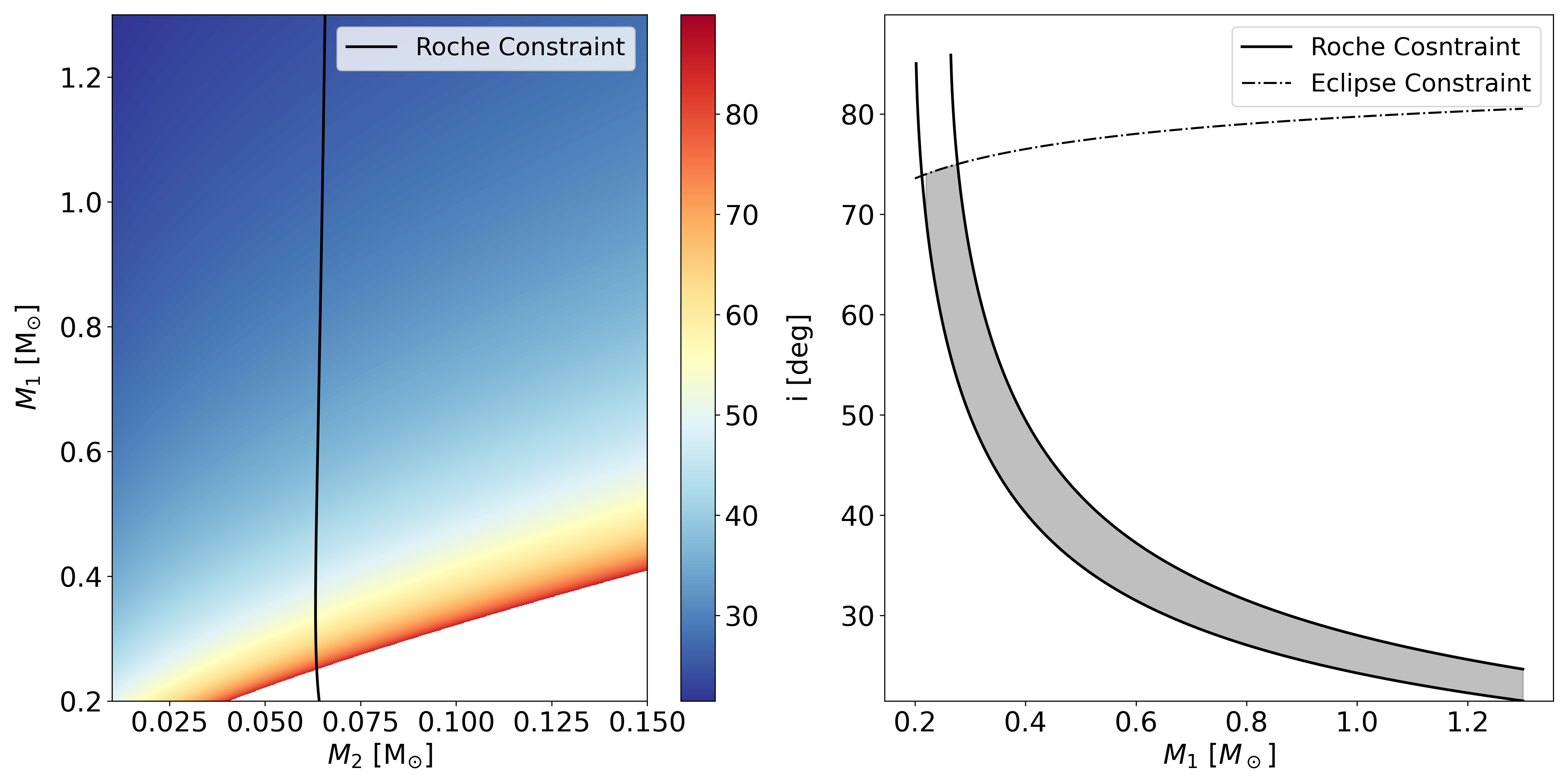}
    \caption{Left: the dynamical constraints on the orbital parameters. The Roche constraint (that the donor must be at the Roche limit for the system), combined with a donor mass-radius relationship yields the nearly vertical black line, i.e., a nearly definite mass for the donor. Each color represents the $M_1-M_2$ constraint from the orbital motion of the narrow component at a fixed inclination. Right: the relationship between the white dwarf mass and system inclination. Given the mass of the donor, the mass of the white dwarf and the inclination are constrained by the kinematic solution to lie on the solid black line. The dot-dashed line marks the upper limit on inclination to avoid a full eclipse of the white dwarf. The allowed parameter space is of the system is shaded in grey. The two curved lines represent the solutions of Equation \ref{eq:binary_mass_func} at the extreme ends of uncertainties in $M_2$, $R_2$ and $K_{RL}$.}
    \label{fig:model_constraints}
\end{figure*}

We use the power-law mass-radius relation from \citet{2011Knigge_CV_evo} for the donors of short-period CVs:
\begin{align}
    \label{eq:knigge_mass_rad_relation}
    \frac{R_2}{R_\odot} = 0.225\pm0.008\left(\frac{M_2}{M_{\textrm{conv}}}\right)^{0.61\pm0.01},
\end{align}
where $M_{\textrm{conv}} = 0.20\pm0.02 M_\odot$ is the donor mass at the period gap (where the donor is fully convective). In contrast to the ordinary brown dwarf mass-radius relation, the relation is constrained by the period-density relation of 
Roche-lobe filling stars on both edges of the period gap \citep{2006Knigge_Period_gap}. Finally, for Keplerian orbits 
\begin{equation}
    a=\left(\frac{P_{\rm orb}^2 G (M_1+M_2)}{4\pi^2}\right)^{1/3}.
    \label{eq:sma}
\end{equation}
Plugging Equations \ref{eq:knigge_mass_rad_relation} and \ref{eq:sma} into Equation \ref{eq:eggleton_roche_lobe}, we obtain one equation for two variables, $M_1$ and $M_2$. 
The locus of the resulting solutions in the $M_1-M_2$ space (solid line in middle panel of Figure \ref{fig:model_constraints}) yields $M_2 = 0.0644\pm 0.0074M_\odot$ which is 
nearly independent of $M_1$. The quoted uncertainty incorporates the stated uncertainties in Equation \ref{eq:knigge_mass_rad_relation}, as well as marginalizing the result over all possible values of $M_1$. This donor mass is on the boundary between low-mass CVs and period bouncers by 
the criteria from \citet{2011Knigge_CV_evo} as expected for the system approximately at the theoretical CV period minimum. 

Our second constraint on the dynamics of the system comes from the radial velocity measurements of the emission lines. The broad component likely has a net streaming velocity and therefore cannot be used for determining the masses of the objects without introducing additional parameters for the motion of the stream. Placing the narrow component at the center of mass of the donor results in a solution with an overly small and statistically unlikely inclination or an unphysically heavy donor. Therefore, we place the emission region producing the narrow component at the Roche point, following \citet{Bonnet1996}, and we assume that the gas is overflowing the Roche point with a negligible net velocity, so the observed velocities are entirely due to the orbital motion at this position. This placement of the narrow component would result in its total or partial blocking by the donor (i.e., dimming) during the phases when the donor is closest to the observer (analogous to HS Cam as deduced by \citealt{Tovmassian1997}), which is in agreement with what we see in Figure \ref{fig:double_peak_lines}. We then use the modified binary mass function to construct a relation between the white dwarf mass $M_1$, donor mass $M_2$ and system orbital inclination $i$ (defined relative to the plane of the sky):
\begin{align}
    \label{eq:binary_mass_func}
    \frac{M_1}{(M_1 + M_2)^{\frac{2}{3}}} = (M_1 + M_2)^{\frac{1}{3}}\frac{R_2}{a}\left(q\right) + \left(\frac{K_{RL}^3P_{\rm orb}}
    {2\pi G\sin^3i}\right)^{\frac{1}{3}}
\end{align}
where $K_{RL}$ is the amplitude of the radial velocity of the narrow component which we assume is located at the Roche point and displays purely orbital motion, and $\frac{R_2}{a}\left(q\right)$ is the Roche lobe radius as a fraction of the semi-major axis from Equation \ref{eq:eggleton_roche_lobe}.

The constraints are visualized in Figure \ref{fig:model_constraints}.
As we do not detect a flat-bottomed photometric dip in the near infrared lightcurve, nor any kind of periodic variability in the optical lightcurve, 
we classify the system as non-eclipsing. Given the size of the donor, this results in an inclination upper limit of $i \sim 75^\circ$, constraining the white dwarf mass to be $M_1\gtrsim 0.262 M_\odot$. Unfortunately, the orbital evolution of the the broad line and of the IR photometric light curve are likely strongly shaped by the magnetic field geometry of the white dwarf, which can have any orientation relative to the orbit, and therefore we cannot use these observations to place any other constraints on the orbital inclination. 
The best estimate of the binary model parameters are listed in Table \ref{tab:star_params}. 

\subsection{The origin of infrared emission and variability}
\label{sec:discussion-ir}

The infrared excess of CSS1603+19 at 1-5\,\micron\ is too bright to be explained by the donor alone. Its short-term variability at the orbital period is surprisingly stable over the 11-year observations from WISE (Fig.~\ref{fig:wise_light_curve}), with a peak-to-peak amplitude of $\sim1$\,mag. While there is no significant difference in the amplitude between W1 and W2, we find a strong wavelength dependence of the infrared variability in the wavelength range of GNIRS spectra between 1-2\,\micron, with the shorter wavelengths having larger amplitude. In addition to the smooth continuum in GNIRS spectra, there are wiggle-like spectral features peaking around 2.14 and 2.39\,\micron\ (Fig.~\ref{fig:cyclotron}). Such short-term periodic variability is only present in infrared but not in optical. CSS1603+19 is among the reddest polars in terms of the cataloged G-W1 color \citep{Ritter2003}. Its 1-mag peak-to-peak amplitude in WISE is also among the largest ones in the polars where their WISE light curves are studied in \cite{Harrison2015}.

We have ruled out the donor as the dominant contributor to the infrared emission. In Sec.~\ref{sec:binary-model}, we show that the donor has a radius of $\sim0.1R_\odot$ according to the observed mass-radius relation of donors \citep{2011Knigge_CV_evo}. If the continuum in GNIRS were due to thermal radiation, its peak at 1.1\,\micron\ in Fig.~\ref{fig:ir_sed_BB_fit} suggests a thermal radiation with a temperature of $\sim2600$\,K, but the observed infrared luminosity is one order of magnitude larger than a $0.1R_\odot$ thermal source. Therefore, the infrared excess is not dominated by the donor. This argument also rules out infrared emission due to the accretion-irradiated donor \citep{Tovmassian1997}: although the time-variability of the infrared emission is consistent with the self-eclipse of the donor's accretor-facing side by the donor itself, the size of required emitting region would still be well above $0.1R_\odot$ and would be incompatible with our dynamical model. The strong magnetic fields of polars prohibit the formation of an accretion disk surrounding the white dwarf, and thus the infrared emission is not from a disk. 

Infrared excess and infrared variability at orbital periods in polars are often associated with cyclotron emission, a byproduct of the accretion shock at the polar caps of the magnetic white dwarf \citep[e.g.][]{Cropper1990,Ferrario1996,Debes2006,Campbell2008a, Campbell2008b, Campbell2008c, Harrison2015}. Cyclotron emission can have multiple harmonic frequencies that span a wide wavelength range, and the wavelengths of the harmonics depend on the strength of magnetic fields. Although the exact wavelengths of the harmonics are uncertain due to the broad spectral profiles of the wiggles, the magnetic field strength needs to be 5-5.5\,MG so that $n=8,9$ harmonics match to the observed wavelengths of 2.14 and 2.39\,\micron, (Sec.~\ref{sec:nir_cont}). With a magnetic field strength of $5$\,MG, its $n=5,6$ cyclotron harmonics would peak at 3.57 and 4.28\,\micron, which may explain the variability in W1 (3.4\,\micron) and W2 (4.5\,\micron).

The origin of the infrared variability at its orbital period remains a mystery. The variability is not due tidal deformation (ellipsoidal variation) of the donor star because ellipsoidal variation would have a variability period at half of the orbital period with an amplitude of $\sim0.3$\,mag at most \citep{Russell1945,McClintock1983}, and the donor flux is insufficient to explain the observed excess for reasonable donor models anyway (Figure \ref{fig:ir_sed_BB_fit}). If the infrared emission is dominated by cyclotron near the surface of the white dwarf, the emitting region may be blocked by the white dwarf itself due to the white dwarf's spin (so-called `self-eclipse'). The self-eclipse scenario can naturally explain the stable infrared periodic variability over 11 years of the WISE data. Since the polar caps are much smaller than the size of the white dwarf, the v-shaped light curves of WISE (Fig.~\ref{fig:wise_light_curve}) require partial self-eclipses of the polar cap; otherwise, a long duration of full self-eclipses would result in a flat feature in the faint phase of the light curve. The polar cap eclipsed by the donor star is unlikely because that would also induce strong variability in optical at the orbital period, inconsistent with our results. 

As seen in Fig. \ref{fig:cyclotron}, the 2.14 and 2.39 \micron\ spectral features are at their flux minimum right at the time when the broad emission-line component is at its maximum redshift (phase $\varphi'=0.69$ in Fig. \ref{fig:cartoon_illustration}). Therefore, it is the back side of the white dwarf -- the one not facing the stream and the donor -- which is producing cyclotron emission. The comparison with emission lines provides key information about the emitting location of the 2.14 and 2.39 \micron\ spectral features. 

While we tentatively identify the 2.14 and 2.39 \micron\ features with cyclotron emission, overall they contribute only a tiny fraction of the infrared flux, with the overall infrared spectrum -- the part of the SED which exhibits strong periodic modulation -- being quite smooth, as seen in Fig. \ref{fig:ir_sed_BB_fit}. Single-zone cyclotron models parametrized by one value of temperature and one value of magnetic field \citep{Chanmugam1981, 1990Schwope_cyclotron_model, Harrison2015} produce well-separated cyclotron peaks, except possibly at the highest temperatures \citep{Kolbin2019}, and do not produce a strong continuum shortward of $n=9-10$ harmonics. Therefore, a single-zone model cannot explain the periodically varying component of our target's SED -- either its overall shape or the low equivalent width of its cyclotron features. It is possible that the emission is made up of components with different temperatures and magnetic fields. If that is the case, the magnetic fields must reach a significantly higher value than the 5MG estimated from the visible harmonics to shift the peak of the emission to lower wavelengths and explain the $1-2$ \micron peak of the emission. It is possible that the observed infrared variability is a more complex interplay involving other structures (e.g. eclipse between accretion streams, hot spots, and/or polar caps). Future X-ray observations may be helpful for understanding the geometry of the polar caps (e.g. \citealt{Heise1985}) to decipher the mysterious infrared variability.


\section{Conclusions}
\label{sec:conclusion}

In this paper we have presented a follow-up study of a cataclysmic variable CSS1603+19, which we categorize as a low-mass AM Herculis star. The white dwarf accretor is likely magnetized, leading to cyclotron emission which dominates emission at mid-infrared wavelengths. We tentatively measure cyclotron harmonics in its near-infrared spectrum, allowing us to estimate the magnetic field strength at $\sim 5 MG$. 

Using NIR and optical spectroscopy, we detect double-peaked emission lines whose velocities and amplitudes vary throughout the orbital period. We build a geometric model of the system (Figure \ref{fig:cartoon_illustration}) in which the broad component originates from the accretion stream near the white dwarf and the narrow component originates from the Roche lobe near the Lagrange point. This model well explains the orbital variability of the lines, and if we further allow for the stream to be bent by the white dwarf's magnetic field away from the direction implied by the orbital dynamics of the gas, it explains the phase offset between the components as well. 

Requiring that the donor must fill its Roche lobe and that it obeys the standard mass-radius relationship for the CV donors, we constrain the mass of the donor to be $M_2=0.0644\pm0.0074 M_\odot$, with an accuracy of about 10\% (this uncertainty is dominated by the uncertainty in the mass-radius relationship we use). The kinematics of the emission lines and the lack of the optical eclipse of the accreting white dwarf then allows us to place constraints on the white dwarf mass ($M_1>0.24 M_{\odot}$) and the inclination ($i\lesssim 75^{\rm o}$). The system is likely caught right before its period bounce. 

\section*{Acknowledgments}

The authors are grateful to S.Littlefair for the constructive and supportive referee report. Y.L., H.-C.H. and N.L.Z. acknowledge support of NASA ADAP grant 80NSSC19K0581 and support from JHU via the seed grant from the Institute for Data Intensive Engineering and Science. H.-C.H. acknowledges the support of the Infosys Membership at the Institute for Advanced Study. N.L.Z. was supported at the IAS by the J. Robert Oppenheimer Visiting Professorship and the Bershadsky Fund.

This work is based on observations obtained at the international Gemini Observatory, a program of NSF’s NOIRLab, which is managed by the Association of Universities for Research in Astronomy (AURA) under a cooperative agreement with the National Science Foundation on behalf of the Gemini Observatory partnership: the National Science Foundation (United States), National Research Council (Canada), Agencia Nacional de Investigaci\'{o}n y Desarrollo (Chile), Ministerio de Ciencia, Tecnolog\'{i}a e Innovaci\'{o}n (Argentina), Minist\'{e}rio da Ci\^{e}ncia, Tecnologia, Inova\c{c}\~{o}es e Comunica\c{c}\~{o}es (Brazil), and Korea Astronomy and Space Science Institute (Republic of Korea).

This publication makes use of data products from the Wide-field Infrared Survey Explorer, which is a joint project of the University of California, Los Angeles, and the Jet Propulsion Laboratory/California Institute of Technology, funded by the National Aeronautics and Space Administration.

This work mades use of data from the European Space Agency (ESA) mission
{\it Gaia} (\url{https://www.cosmos.esa.int/gaia}), processed by the {\it Gaia}
Data Processing and Analysis Consortium (DPAC,
\url{https://www.cosmos.esa.int/web/gaia/dpac/consortium}). Funding for the DPAC
has been provided by national institutions, in particular the institutions
participating in the {\it Gaia} Multilateral Agreement.

This work is based on observations obtained with the Samuel Oschin Telescope 48-inch and the 60-inch Telescope at the Palomar
Observatory as part of the Zwicky Transient Facility project. ZTF is supported by the National Science Foundation under Grants
No. AST-1440341 and AST-2034437 and a collaboration including current partners Caltech, IPAC, the Weizmann Institute for
Science, the Oskar Klein Center at Stockholm University, the University of Maryland, Deutsches Elektronen-Synchrotron and
Humboldt University, the TANGO Consortium of Taiwan, the University of Wisconsin at Milwaukee, Trinity College Dublin,
Lawrence Livermore National Laboratories, IN2P3, University of Warwick, Ruhr University Bochum, Northwestern University and
former partners the University of Washington, Los Alamos National Laboratories, and Lawrence Berkeley National Laboratories.
Operations are conducted by COO, IPAC, and UW.

This work is based on observations obtained at the MDM Observatory, operated by Dartmouth College, Columbia University, Ohio State University, 
Ohio University, and the University of Michigan.

This work is based on observations obtained with the Apache Point Observatory 3.5-meter telescope, which is owned and operated 
by the Astrophysical Research Consortium.

This research has made use of the NASA/IPAC Infrared Science Archive, which is funded by the National Aeronautics and Space Administration and operated by the California Institute of Technology.

\section{Data Availability}
The WISE and ZTF data sets are publicly available at in NASA/IPAC Infrared Science Archive (\href{https://irsa.ipac.caltech.edu/frontpage/}{https://irsa.ipac.caltech.edu/frontpage/}).
The Gaia DR3 data sets are publicly available in the Gaia Archive at ESA (\href{https://gea.esac.esa.int/archive/}{https://gea.esac.esa.int/archive/}).
The GNIRS, MDM and APO spectroscopy data underlying this article will be shared on reasonable request to the corresponding author.

\bibliographystyle{mnras}
\bibliography{CV1603+19}

\begin{thebibliography}{}
\makeatletter
\relax
\def\mn@urlcharsother{\let\do\@makeother \do\$\do\&\do\#\do\^\do\_\do\%\do\~}
\def\mn@doi{\begingroup\mn@urlcharsother \@ifnextchar [ {\mn@doi@}
  {\mn@doi@[]}}
\def\mn@doi@[#1]#2{\def\@tempa{#1}\ifx\@tempa\@empty \href
  {http://dx.doi.org/#2} {doi:#2}\else \href {http://dx.doi.org/#2} {#1}\fi
  \endgroup}
\def\mn@eprint#1#2{\mn@eprint@#1:#2::\@nil}
\def\mn@eprint@arXiv#1{\href {http://arxiv.org/abs/#1} {{\tt arXiv:#1}}}
\def\mn@eprint@dblp#1{\href {http://dblp.uni-trier.de/rec/bibtex/#1.xml}
  {dblp:#1}}
\def\mn@eprint@#1:#2:#3:#4\@nil{\def\@tempa {#1}\def\@tempb {#2}\def\@tempc
  {#3}\ifx \@tempc \@empty \let \@tempc \@tempb \let \@tempb \@tempa \fi \ifx
  \@tempb \@empty \def\@tempb {arXiv}\fi \@ifundefined
  {mn@eprint@\@tempb}{\@tempb:\@tempc}{\expandafter \expandafter \csname
  mn@eprint@\@tempb\endcsname \expandafter{\@tempc}}}

\bibitem[\protect\citeauthoryear{{Astropy Collaboration} et~al.,}{{Astropy
  Collaboration} et~al.}{2018}]{2018astropy}
{Astropy Collaboration} et~al., 2018, \mn@doi [\aj] {10.3847/1538-3881/aabc4f},
  \href {https://ui.adsabs.harvard.edu/abs/2018AJ....156..123A} {156, 123}

\bibitem[\protect\citeauthoryear{{Bonnet-Bidaud}, {Mouchet}, {Somova}  \&
  {Somov}}{{Bonnet-Bidaud} et~al.}{1996}]{Bonnet1996}
{Bonnet-Bidaud} J.~M.,  {Mouchet} M.,  {Somova} T.~A.,   {Somov} N.~N.,  1996,
  \aap, \href {https://ui.adsabs.harvard.edu/abs/1996A&A...306..199B} {306,
  199}

\bibitem[\protect\citeauthoryear{{Campbell}, {Harrison}, {Schwope}  \&
  {Howell}}{{Campbell} et~al.}{2008a}]{Campbell2008a}
{Campbell} R.~K.,  {Harrison} T.~E.,  {Schwope} A.~D.,   {Howell} S.~B.,
  2008a, \mn@doi [\apj] {10.1086/523632}, \href
  {https://ui.adsabs.harvard.edu/abs/2008ApJ...672..531C} {672, 531}

\bibitem[\protect\citeauthoryear{{Campbell}, {Harrison}, {Mason}, {Howell}  \&
  {Schwope}}{{Campbell} et~al.}{2008b}]{Campbell2008b}
{Campbell} R.~K.,  {Harrison} T.~E.,  {Mason} E.,  {Howell} S.,   {Schwope}
  A.~D.,  2008b, \mn@doi [\apj] {10.1086/533488}, \href
  {https://ui.adsabs.harvard.edu/abs/2008ApJ...678.1304C} {678, 1304}

\bibitem[\protect\citeauthoryear{{Campbell}, {Harrison}  \& {Kafka}}{{Campbell}
  et~al.}{2008c}]{Campbell2008c}
{Campbell} R.~K.,  {Harrison} T.~E.,   {Kafka} S.,  2008c, \mn@doi [\apj]
  {10.1086/589179}, \href
  {https://ui.adsabs.harvard.edu/abs/2008ApJ...683..409C} {683, 409}

\bibitem[\protect\citeauthoryear{{Chanmugam} \& {Dulk}}{{Chanmugam} \&
  {Dulk}}{1981}]{Chanmugam1981}
{Chanmugam} G.,  {Dulk} G.~A.,  1981, \mn@doi [\apj] {10.1086/158736}, \href
  {https://ui.adsabs.harvard.edu/abs/1981ApJ...244..569C} {244, 569}

\bibitem[\protect\citeauthoryear{{Ciardi}, {Howell}, {Hauschildt}  \&
  {Allard}}{{Ciardi} et~al.}{1998}]{Ciardi1998}
{Ciardi} D.~R.,  {Howell} S.~B.,  {Hauschildt} P.~H.,   {Allard} F.,  1998,
  \mn@doi [\apj] {10.1086/306081}, \href
  {https://ui.adsabs.harvard.edu/abs/1998ApJ...504..450C} {504, 450}

\bibitem[\protect\citeauthoryear{{Cropper}}{{Cropper}}{1990}]{1990Cropper_Polar_Review}
{Cropper} M.,  1990, \mn@doi [\ssr] {10.1007/BF00177799}, \href
  {https://ui.adsabs.harvard.edu/abs/1990SSRv...54..195C} {54, 195}

\bibitem[\protect\citeauthoryear{{Cropper} et~al.,}{{Cropper}
  et~al.}{1990}]{Cropper1990}
{Cropper} M.,  et~al., 1990, \mnras, \href
  {https://ui.adsabs.harvard.edu/abs/1990MNRAS.245..760C} {245, 760}

\bibitem[\protect\citeauthoryear{{Cutri} et~al.,}{{Cutri}
  et~al.}{2012}]{WISE_DATA_GUILD_2012}
{Cutri} R.~M.,  et~al., 2012, {Explanatory Supplement to the WISE All-Sky Data
  Release Products}, Explanatory Supplement to the WISE All-Sky Data Release
  Products

\bibitem[\protect\citeauthoryear{{Debes}, {L{\'o}pez-Morales}, {Bonanos}  \&
  {Weinberger}}{{Debes} et~al.}{2006}]{Debes2006}
{Debes} J.~H.,  {L{\'o}pez-Morales} M.,  {Bonanos} A.~Z.,   {Weinberger} A.~J.,
   2006, \mn@doi [\apjl] {10.1086/507486}, \href
  {https://ui.adsabs.harvard.edu/abs/2006ApJ...647L.147D} {647, L147}

\bibitem[\protect\citeauthoryear{{Dekany} et~al.,}{{Dekany}
  et~al.}{2020}]{2020ZTF_facility}
{Dekany} R.,  et~al., 2020, \mn@doi [\pasp] {10.1088/1538-3873/ab4ca2}, \href
  {https://ui.adsabs.harvard.edu/abs/2020PASP..132c8001D} {132, 038001}

\bibitem[\protect\citeauthoryear{{Dubus}, {Campbell}, {Kern}, {Taam}  \&
  {Spruit}}{{Dubus} et~al.}{2004}]{2004Dubus_mid_IR_excess}
{Dubus} G.,  {Campbell} R.,  {Kern} B.,  {Taam} R.~E.,   {Spruit} H.~C.,  2004,
  \mn@doi [\mnras] {10.1111/j.1365-2966.2004.07551.x}, \href
  {https://ui.adsabs.harvard.edu/abs/2004MNRAS.349..869D} {349, 869}

\bibitem[\protect\citeauthoryear{{Eggleton}}{{Eggleton}}{1983}]{1983Eggleton_Roche_radius}
{Eggleton} P.~P.,  1983, \mn@doi [\apj] {10.1086/160960}, \href
  {https://ui.adsabs.harvard.edu/abs/1983ApJ...268..368E} {268, 368}

\bibitem[\protect\citeauthoryear{{Elias}, {Rodgers}, {Joyce}, {Lazo},
  {Doppmann}, {Winge}  \& {Rodr{\'\i}guez-Ardila}}{{Elias}
  et~al.}{2006a}]{2006GNIRS_Spec_performance}
{Elias} J.~H.,  {Rodgers} B.,  {Joyce} R.~R.,  {Lazo} M.,  {Doppmann} G.,
  {Winge} C.,   {Rodr{\'\i}guez-Ardila} A.,  2006a, in {McLean} I.~S.,  {Iye}
  M.,  eds,  Society of Photo-Optical Instrumentation Engineers (SPIE)
  Conference Series Vol. 6269, Society of Photo-Optical Instrumentation
  Engineers (SPIE) Conference Series. p. 626914, \mn@doi{10.1117/12.671765}

\bibitem[\protect\citeauthoryear{{Elias}, {Joyce}, {Liang}, {Muller}, {Hileman}
   \& {George}}{{Elias} et~al.}{2006b}]{GNIRS_Design}
{Elias} J.~H.,  {Joyce} R.~R.,  {Liang} M.,  {Muller} G.~P.,  {Hileman} E.~A.,
   {George} J.~R.,  2006b, in {McLean} I.~S.,  {Iye} M.,  eds,  Society of
  Photo-Optical Instrumentation Engineers (SPIE) Conference Series Vol. 6269,
  Society of Photo-Optical Instrumentation Engineers (SPIE) Conference Series.
  p. 62694C, \mn@doi{10.1117/12.671817}

\bibitem[\protect\citeauthoryear{{Ferrario}, {Bailey}  \&
  {Wickramasinghe}}{{Ferrario} et~al.}{1996}]{Ferrario1996}
{Ferrario} L.,  {Bailey} J.,   {Wickramasinghe} D.,  1996, \mn@doi [\mnras]
  {10.1093/mnras/282.1.218}, \href
  {https://ui.adsabs.harvard.edu/abs/1996MNRAS.282..218F} {282, 218}

\bibitem[\protect\citeauthoryear{{Gaia Collaboration} et~al.,}{{Gaia
  Collaboration} et~al.}{2016}]{2016Gaia_mission}
{Gaia Collaboration} et~al., 2016, \mn@doi [\aap]
  {10.1051/0004-6361/201629272}, \href
  {https://ui.adsabs.harvard.edu/abs/2016A&A...595A...1G} {595, A1}

\bibitem[\protect\citeauthoryear{{Gaia Collaboration} et~al.,}{{Gaia
  Collaboration} et~al.}{2022}]{GaiaDR3Vallenari2022}
{Gaia Collaboration} et~al., 2022, arXiv e-prints, \href
  {https://ui.adsabs.harvard.edu/abs/2022arXiv220800211G} {p. arXiv:2208.00211}

\bibitem[\protect\citeauthoryear{{G{\"a}nsicke} et~al.,}{{G{\"a}nsicke}
  et~al.}{2009}]{Gaensicke2009}
{G{\"a}nsicke} B.~T.,  et~al., 2009, \mn@doi [\mnras]
  {10.1111/j.1365-2966.2009.15126.x}, \href
  {https://ui.adsabs.harvard.edu/abs/2009MNRAS.397.2170G} {397, 2170}

\bibitem[\protect\citeauthoryear{{Hameury}, {King}, {Lasota}  \&
  {Ritter}}{{Hameury} et~al.}{1988}]{1988Hameury_magnetic_braking}
{Hameury} J.~M.,  {King} A.~R.,  {Lasota} J.~P.,   {Ritter} H.,  1988, \mn@doi
  [\mnras] {10.1093/mnras/231.3.535}, \href
  {https://ui.adsabs.harvard.edu/abs/1988MNRAS.231..535H} {231, 535}

\bibitem[\protect\citeauthoryear{{Harrison} \& {Campbell}}{{Harrison} \&
  {Campbell}}{2015}]{Harrison2015}
{Harrison} T.~E.,  {Campbell} R.~K.,  2015, \mn@doi [\apjs]
  {10.1088/0067-0049/219/2/32}, \href
  {https://ui.adsabs.harvard.edu/abs/2015ApJS..219...32H} {219, 32}

\bibitem[\protect\citeauthoryear{{Harrison}, {Hamilton}, {Tappert}, {Hoffman}
  \& {Campbell}}{{Harrison} et~al.}{2013}]{Harrison2013}
{Harrison} T.~E.,  {Hamilton} R.~T.,  {Tappert} C.,  {Hoffman} D.~I.,
  {Campbell} R.~K.,  2013, \mn@doi [\aj] {10.1088/0004-6256/145/1/19}, \href
  {https://ui.adsabs.harvard.edu/abs/2013AJ....145...19H} {145, 19}

\bibitem[\protect\citeauthoryear{{Heise}, {Brinkman}, {Gronenschild}, {Watson},
  {King}, {Stella}  \& {Kieboom}}{{Heise} et~al.}{1985}]{Heise1985}
{Heise} J.,  {Brinkman} A.~C.,  {Gronenschild} E.,  {Watson} M.,  {King} A.~R.,
   {Stella} L.,   {Kieboom} K.,  1985, \aap, \href
  {https://ui.adsabs.harvard.edu/abs/1985A&A...148L..14H} {148, L14}

\bibitem[\protect\citeauthoryear{{Hertzsprung}}{{Hertzsprung}}{1909}]{1909HR_diagram}
{Hertzsprung} E.,  1909, \mn@doi [Astronomische Nachrichten]
  {10.1002/asna.19081792402}, \href
  {https://ui.adsabs.harvard.edu/abs/1909AN....179..373H} {179, 373}

\bibitem[\protect\citeauthoryear{{Horne}}{{Horne}}{1986}]{1986PASP_Horne_Spec_ex_algo}
{Horne} K.,  1986, \mn@doi [\pasp] {10.1086/131801}, \href
  {https://ui.adsabs.harvard.edu/abs/1986PASP...98..609H} {98, 609}

\bibitem[\protect\citeauthoryear{{Hwang} \& {Zakamska}}{{Hwang} \&
  {Zakamska}}{2020}]{Hwang_2020_EBs}
{Hwang} H.-C.,  {Zakamska} N.~L.,  2020, \mn@doi [\mnras]
  {10.1093/mnras/staa400}, \href
  {https://ui.adsabs.harvard.edu/abs/2020MNRAS.493.2271H} {493, 2271}

\bibitem[\protect\citeauthoryear{{Kalomeni}, {Nelson}, {Rappaport}, {Molnar},
  {Quintin}  \& {Yakut}}{{Kalomeni} et~al.}{2016}]{2016Kalomeni_CV_evo}
{Kalomeni} B.,  {Nelson} L.,  {Rappaport} S.,  {Molnar} M.,  {Quintin} J.,
  {Yakut} K.,  2016, \mn@doi [\apj] {10.3847/1538-4357/833/1/83}, \href
  {https://ui.adsabs.harvard.edu/abs/2016ApJ...833...83K} {833, 83}

\bibitem[\protect\citeauthoryear{{Knigge}}{{Knigge}}{2006}]{2006Knigge_Period_gap}
{Knigge} C.,  2006, \mn@doi [\mnras] {10.1111/j.1365-2966.2006.11096.x}, \href
  {https://ui.adsabs.harvard.edu/abs/2006MNRAS.373..484K} {373, 484}

\bibitem[\protect\citeauthoryear{{Knigge}, {Baraffe}  \& {Patterson}}{{Knigge}
  et~al.}{2011}]{2011Knigge_CV_evo}
{Knigge} C.,  {Baraffe} I.,   {Patterson} J.,  2011, \mn@doi [\apjs]
  {10.1088/0067-0049/194/2/28}, \href
  {https://ui.adsabs.harvard.edu/abs/2011ApJS..194...28K} {194, 28}

\bibitem[\protect\citeauthoryear{{Koester}}{{Koester}}{2010}]{2010Koester_wd_atm_model}
{Koester} D.,  2010, \memsai, \href
  {https://ui.adsabs.harvard.edu/abs/2010MmSAI..81..921K} {81, 921}

\bibitem[\protect\citeauthoryear{{Kolb}}{{Kolb}}{1993}]{1993Kolb_cv_P_min}
{Kolb} U.,  1993, \aap, \href
  {https://ui.adsabs.harvard.edu/abs/1993A&A...271..149K} {271, 149}

\bibitem[\protect\citeauthoryear{Kolb \& Baraffe}{Kolb \&
  Baraffe}{1999}]{1999Kolb_cv_P_min}
Kolb U.,  Baraffe I.,  1999, Monthly Notices of the Royal Astronomical Society,
  309, 1034

\bibitem[\protect\citeauthoryear{{Kolbin}, {Serebryakova}, {Gabdeev}  \&
  {Borisov}}{{Kolbin} et~al.}{2019}]{Kolbin2019}
{Kolbin} A.~I.,  {Serebryakova} N.~A.,  {Gabdeev} M.~M.,   {Borisov} N.~V.,
  2019, \mn@doi [Astrophysical Bulletin] {10.1134/S1990341319010085}, \href
  {https://ui.adsabs.harvard.edu/abs/2019AstBu..74...80K} {74, 80}

\bibitem[\protect\citeauthoryear{{Lomb}}{{Lomb}}{1976}]{Lomb1976}
{Lomb} N.~R.,  1976, \mn@doi [\apss] {10.1007/BF00648343}, \href
  {https://ui.adsabs.harvard.edu/abs/1976Ap&SS..39..447L} {39, 447}

\bibitem[\protect\citeauthoryear{{Lubow} \& {Shu}}{{Lubow} \&
  {Shu}}{1975}]{1975Lubow_accretion_stream}
{Lubow} S.~H.,  {Shu} F.~H.,  1975, \mn@doi [\apj] {10.1086/153614}, \href
  {https://ui.adsabs.harvard.edu/abs/1975ApJ...198..383L} {198, 383}

\bibitem[\protect\citeauthoryear{{Mainzer} et~al.,}{{Mainzer}
  et~al.}{2011}]{Mainzer_2011}
{Mainzer} A.,  et~al., 2011, \mn@doi [\apj] {10.1088/0004-637X/731/1/53}, \href
  {https://ui.adsabs.harvard.edu/abs/2011ApJ...731...53M} {731, 53}

\bibitem[\protect\citeauthoryear{{Marley} et~al.,}{{Marley}
  et~al.}{2021}]{2021Marley_BD_template}
{Marley} M.~S.,  et~al., 2021, \mn@doi [\apj] {10.3847/1538-4357/ac141d}, \href
  {https://ui.adsabs.harvard.edu/abs/2021ApJ...920...85M} {920, 85}

\bibitem[\protect\citeauthoryear{Masci et~al.,}{Masci
  et~al.}{2018}]{2019Masci_ZTF_Description}
Masci F.~J.,  et~al., 2018, \mn@doi [Publications of the Astronomical Society
  of the Pacific] {10.1088/1538-3873/aae8ac}, 131, 018003

\bibitem[\protect\citeauthoryear{{Mason}, {Wells}, {Motsoaledi}, {Szkody}  \&
  {Gonzalez}}{{Mason} et~al.}{2019}]{Mason2019}
{Mason} P.~A.,  {Wells} N.~K.,  {Motsoaledi} M.,  {Szkody} P.,   {Gonzalez} E.,
   2019, \mn@doi [\mnras] {10.1093/mnras/stz1863}, \href
  {https://ui.adsabs.harvard.edu/abs/2019MNRAS.488.2881M} {488, 2881}

\bibitem[\protect\citeauthoryear{{McClintock}, {Petro}, {Remillard}  \&
  {Ricker}}{{McClintock} et~al.}{1983}]{McClintock1983}
{McClintock} J.~E.,  {Petro} L.~D.,  {Remillard} R.~A.,   {Ricker} G.~R.,
  1983, \mn@doi [\apjl] {10.1086/183972}, \href
  {https://ui.adsabs.harvard.edu/abs/1983ApJ...266L..27M} {266, L27}

\bibitem[\protect\citeauthoryear{{Mestel} \& {Spruit}}{{Mestel} \&
  {Spruit}}{1987}]{1987Mestel_magnetic_braking}
{Mestel} L.,  {Spruit} H.~C.,  1987, \mn@doi [\mnras] {10.1093/mnras/226.1.57},
  \href {https://ui.adsabs.harvard.edu/abs/1987MNRAS.226...57M} {226, 57}

\bibitem[\protect\citeauthoryear{{Oliveira}, {Rodrigues}, {Martins},
  {Palhares}, {Silva}, {Lima}  \& {Jablonski}}{{Oliveira}
  et~al.}{2020}]{Oliveira2020}
{Oliveira} A.~S.,  {Rodrigues} C.~V.,  {Martins} M.,  {Palhares} M.~S.,
  {Silva} K.~M.~G.,  {Lima} I.~J.,   {Jablonski} F.~J.,  2020, \mn@doi [\aj]
  {10.3847/1538-3881/ab6ded}, \href
  {https://ui.adsabs.harvard.edu/abs/2020AJ....159..114O} {159, 114}

\bibitem[\protect\citeauthoryear{{Paczynski} \& {Sienkiewicz}}{{Paczynski} \&
  {Sienkiewicz}}{1981}]{Paczynski1981}
{Paczynski} B.,  {Sienkiewicz} R.,  1981, \mn@doi [\apjl] {10.1086/183616},
  \href {https://ui.adsabs.harvard.edu/abs/1981ApJ...248L..27P} {248, L27}

\bibitem[\protect\citeauthoryear{{Patterson}}{{Patterson}}{1984}]{Patterson1984}
{Patterson} J.,  1984, \mn@doi [\apjs] {10.1086/190940}, \href
  {https://ui.adsabs.harvard.edu/abs/1984ApJS...54..443P} {54, 443}

\bibitem[\protect\citeauthoryear{{Peterson} et~al.,}{{Peterson}
  et~al.}{2008}]{2008Peterson_Brown_dwarf_atmosphere}
{Peterson} D.~E.,  et~al., 2008, \mn@doi [\apj] {10.1086/590527}, \href
  {https://ui.adsabs.harvard.edu/abs/2008ApJ...685..313P} {685, 313}

\bibitem[\protect\citeauthoryear{{Petrosky}, {Hwang}, {Zakamska}, {Chandra}  \&
  {Hill}}{{Petrosky} et~al.}{2021}]{Petrosky_2021}
{Petrosky} E.,  {Hwang} H.-C.,  {Zakamska} N.~L.,  {Chandra} V.,   {Hill}
  M.~J.,  2021, \mn@doi [\mnras] {10.1093/mnras/stab592}, \href
  {https://ui.adsabs.harvard.edu/abs/2021MNRAS.503.3975P} {503, 3975}

\bibitem[\protect\citeauthoryear{Prochaska et~al.,}{Prochaska
  et~al.}{2020}]{Prochaska2020Pypeit}
Prochaska J.~X.,  et~al., 2020, \mn@doi [Journal of Open Source Software]
  {10.21105/joss.02308}, 5, 2308

\bibitem[\protect\citeauthoryear{{Rappaport}, {Joss}  \& {Webbink}}{{Rappaport}
  et~al.}{1982}]{1982Rappaport_compact_binary_evo}
{Rappaport} S.,  {Joss} P.~C.,   {Webbink} R.~F.,  1982, \mn@doi [\apj]
  {10.1086/159772}, \href
  {https://ui.adsabs.harvard.edu/abs/1982ApJ...254..616R} {254, 616}

\bibitem[\protect\citeauthoryear{{Rappaport}, {Verbunt}  \& {Joss}}{{Rappaport}
  et~al.}{1983}]{Rappaport1983}
{Rappaport} S.,  {Verbunt} F.,   {Joss} P.~C.,  1983, \mn@doi [\apj]
  {10.1086/161569}, \href
  {https://ui.adsabs.harvard.edu/abs/1983ApJ...275..713R} {275, 713}

\bibitem[\protect\citeauthoryear{{Ritter} \& {Kolb}}{{Ritter} \&
  {Kolb}}{2003}]{Ritter2003}
{Ritter} H.,  {Kolb} U.,  2003, \mn@doi [\aap] {10.1051/0004-6361:20030330},
  \href {https://ui.adsabs.harvard.edu/abs/2003A&A...404..301R} {404, 301}

\bibitem[\protect\citeauthoryear{{Rosenberg}}{{Rosenberg}}{1910}]{1910HR_diagram}
{Rosenberg} H.,  1910, \mn@doi [Astronomische Nachrichten]
  {10.1002/asna.19101860503}, \href
  {https://ui.adsabs.harvard.edu/abs/1910AN....186...71R} {186, 71}

\bibitem[\protect\citeauthoryear{{Russell}}{{Russell}}{1914}]{Russell_1914}
{Russell} H.~N.,  1914, Popular Astronomy, \href
  {https://ui.adsabs.harvard.edu/abs/1914PA.....22..275R} {22, 275}

\bibitem[\protect\citeauthoryear{{Russell}}{{Russell}}{1945}]{Russell1945}
{Russell} H.~N.,  1945, \mn@doi [\apj] {10.1086/144733}, \href
  {https://ui.adsabs.harvard.edu/abs/1945ApJ...102....1R} {102, 1}

\bibitem[\protect\citeauthoryear{{Scargle}}{{Scargle}}{1982}]{Scargle1982}
{Scargle} J.~D.,  1982, \mn@doi [\apj] {10.1086/160554}, \href
  {https://ui.adsabs.harvard.edu/abs/1982ApJ...263..835S} {263, 835}

\bibitem[\protect\citeauthoryear{{Schwope}}{{Schwope}}{1990}]{1990Schwope_cyclotron_model}
{Schwope} A.~D.,  1990, \mn@doi [Reviews in Modern Astronomy]
  {10.1007/978-3-642-76238-3_5}, \href
  {https://ui.adsabs.harvard.edu/abs/1990RvMA....3...44S} {3, 44}

\bibitem[\protect\citeauthoryear{{Schwope} \& {Beuermann}}{{Schwope} \&
  {Beuermann}}{1990}]{Schwope_1990_V834}
{Schwope} A.~D.,  {Beuermann} K.,  1990, \aap, \href
  {https://ui.adsabs.harvard.edu/abs/1990A&A...238..173S} {238, 173}

\bibitem[\protect\citeauthoryear{{Schwope}, {Beuermann}, {Jordan}  \&
  {Thomas}}{{Schwope} et~al.}{1993}]{Schwope_1993_MRSerp}
{Schwope} A.~D.,  {Beuermann} K.,  {Jordan} S.,   {Thomas} H.~C.,  1993, \aap,
  \href {https://ui.adsabs.harvard.edu/abs/1993A&A...278..487S} {278, 487}

\bibitem[\protect\citeauthoryear{{Silber}}{{Silber}}{1992}]{Silber1992}
{Silber} A.~D.,  1992, PhD thesis, Massachusetts Institute of Technology

\bibitem[\protect\citeauthoryear{{Sirotkin} \& {Kim}}{{Sirotkin} \&
  {Kim}}{2009}]{2009Sirotkin_Roche_approx}
{Sirotkin} F.~V.,  {Kim} W.-T.,  2009, \mn@doi [\apj]
  {10.1088/0004-637X/698/1/715}, \href
  {https://ui.adsabs.harvard.edu/abs/2009ApJ...698..715S} {698, 715}

\bibitem[\protect\citeauthoryear{{Spruit} \& {Ritter}}{{Spruit} \&
  {Ritter}}{1983}]{1983Spruit_CV_period_gap}
{Spruit} H.~C.,  {Ritter} H.,  1983, \aap, \href
  {https://ui.adsabs.harvard.edu/abs/1983A&A...124..267S} {124, 267}

\bibitem[\protect\citeauthoryear{{Taam} \& {Spruit}}{{Taam} \&
  {Spruit}}{1989}]{1989Taam_magnetic_braking}
{Taam} R.~E.,  {Spruit} H.~C.,  1989, \mn@doi [\apj] {10.1086/167966}, \href
  {https://ui.adsabs.harvard.edu/abs/1989ApJ...345..972T} {345, 972}

\bibitem[\protect\citeauthoryear{Thorstensen \& Skinner}{Thorstensen \&
  Skinner}{2012}]{Thorstensen_2012cvperiods}
Thorstensen J.~R.,  Skinner J.~N.,  2012, \mn@doi [The Astronomical Journal]
  {10.1088/0004-6256/144/3/81}, 144, 81

\bibitem[\protect\citeauthoryear{Thorstensen, Ringwald, Taylor, Sheets, Peters,
  Skinner, Alper  \& Weil}{Thorstensen
  et~al.}{2017}]{Thorstensen_2017cv_period_note}
Thorstensen J.~R.,  Ringwald F.~A.,  Taylor C.~J.,  Sheets H.~A.,  Peters
  C.~S.,  Skinner J.~N.,  Alper E.~H.,   Weil K.~E.,  2017, \mn@doi [Research
  Notes of the {AAS}] {10.3847/2515-5172/aa9d2a}, 1, 29

\bibitem[\protect\citeauthoryear{{Tovmassian}, {Greiner}, {Zickgraf}, {Kroll},
  {Krautter}, {Thiering}, {Zharykov}  \& {Serrano}}{{Tovmassian}
  et~al.}{1997}]{Tovmassian1997}
{Tovmassian} G.~H.,  {Greiner} J.,  {Zickgraf} F.~J.,  {Kroll} P.,  {Krautter}
  J.,  {Thiering} I.,  {Zharykov} S.~V.,   {Serrano} A.,  1997, \aap, \href
  {https://ui.adsabs.harvard.edu/abs/1997A&A...328..571T} {328, 571}

\bibitem[\protect\citeauthoryear{{Tremblay} \& {Bergeron}}{{Tremblay} \&
  {Bergeron}}{2009}]{2009Tremblay_wd_atm_model}
{Tremblay} P.~E.,  {Bergeron} P.,  2009, \mn@doi [\apj]
  {10.1088/0004-637X/696/2/1755}, \href
  {https://ui.adsabs.harvard.edu/abs/2009ApJ...696.1755T} {696, 1755}

\bibitem[\protect\citeauthoryear{{Verbunt}}{{Verbunt}}{1984}]{1984Verbunt_mass_transfer_and_period_gap}
{Verbunt} F.,  1984, \mn@doi [\mnras] {10.1093/mnras/209.2.227}, \href
  {https://ui.adsabs.harvard.edu/abs/1984MNRAS.209..227V} {209, 227}

\bibitem[\protect\citeauthoryear{{Verbunt} \& {Zwaan}}{{Verbunt} \&
  {Zwaan}}{1981}]{1981Verbunt_MB}
{Verbunt} F.,  {Zwaan} C.,  1981, \aap, \href
  {https://ui.adsabs.harvard.edu/abs/1981A&A...100L...7V} {100, L7}

\bibitem[\protect\citeauthoryear{Warner}{Warner}{1995}]{Warner1995}
Warner B.,  1995, Cataclysmic Variable Stars.
Cambridge Astrophysics, Cambridge University Press,
  \mn@doi{10.1017/CBO9780511586491}

\bibitem[\protect\citeauthoryear{{Wright} et~al.,}{{Wright}
  et~al.}{2010}]{WISE_Spec_2010}
{Wright} E.~L.,  et~al., 2010, \mn@doi [\aj] {10.1088/0004-6256/140/6/1868},
  \href {https://ui.adsabs.harvard.edu/abs/2010AJ....140.1868W} {140, 1868}

\bibitem[\protect\citeauthoryear{{van Paradijs}}{{van
  Paradijs}}{1986}]{1986Paradijs_MB}
{van Paradijs} J.,  1986, \mn@doi [\mnras] {10.1093/mnras/218.1.31P}, \href
  {https://ui.adsabs.harvard.edu/abs/1986MNRAS.218P..31V} {218, 31p}

\makeatother
\end{thebibliography}

\label{lastpage}
\end{document}